\documentclass{article}
\usepackage{graphicx}  
\usepackage{amsmath}   
\usepackage[compress]{cite}
\usepackage{amssymb}   
\usepackage{bm} 
\usepackage{dcolumn}
\usepackage{cancel}
\usepackage{color}
\usepackage{mathrsfs}
\usepackage{amsfonts}
\usepackage{varioref}
\usepackage{textcomp}
\usepackage{subcaption}
\usepackage{comment}
\RequirePackage[colorlinks,citecolor=blue,urlcolor=magenta,linkcolor=blue]{hyperref}
\allowdisplaybreaks
\newcommand{\eps}{\epsilon}

\newcommand{\p}{\partial}

\def\mbh{M_{\rm BH}}

\def\drm{\mathrm{d}} 
\def\order{\mathcal{O}}
\labelformat{section}{Section #1} 
\labelformat{subsection}{Section #1} 
\labelformat{subsubsection}{Section #1}
\labelformat{subsubsubsection}{Section #1}
\labelformat{figure}{Figure~#1} 
\labelformat{subfigure}{Fig.~\thefigure#1} 
\labelformat{table}{Table~#1} 
\labelformat{appendix}{Appendix #1}
\title
{\bf Tidal Love numbers and quasi-normal modes of the Schwarzschild-Hernquist black hole}
\author{Sumanta Chakraborty\footnote{tpsc@iacs.res.in}$~^{1}$, Geoffrey Comp\`ere\footnote{geoffrey.compere@ulb.be}$~^{2}$ and Ludovico Machet\footnote{ludovico.machet@kuleuven.be}$~^{2,3,4}$
\\
$^{1}${\small{School of Physical Sciences, Indian Association for the Cultivation of Science, Kolkata-700032, India}}\\
\centerline{$^{2}$\small{Universit\'{e} Libre de Bruxelles, BLU-ULB space center, International Solvay Institutes,}}\\
\centerline{\small{CP 231, B-1050 Brussels, Belgium}}\\
\centerline{$^{3}$\small{Institute for Theoretical Physics, KU Leuven, Celestijnenlaan 200D, B-3001 Leuven, Belgium}}\\
\centerline{$^{4}$\small{Leuven Gravity Institute, KU Leuven, Celestijnenlaan 200D, B-3001 Leuven, Belgium}}
}

\begin{document}
  
\maketitle
\begin{abstract}
We derive the model of the Schwarzschild black hole immersed into a dark matter halo with a relativistic Hernquist profile, the Schwarzschild-Hernquist black hole, and obtain its tidal Love numbers and quasi-normal modes. We thoroughly compare our odd and even parity perturbation equations with the literature and point out that two distinct choices of matter perturbations lead to distinct spectra. We establish that the quasi-normal modes admit qualitatively distinct scaling laws in terms of dark matter densities for non-relativistic and relativistic halos. We develop a stable numerical scheme for computing tidal Love numbers based on asymptotic series expansions. We further comment upon the existence of matter configurations obeying the dominant energy condition that lead to multiple light rings. 
\end{abstract}

\tableofcontents

\section{Introduction}

The detections of merging compact objects, primarily black holes (BH) and neutron stars, via gravitational waves (GW) opened a new window on the universe and kicked off the phenomenological study of gravity in its strongly dynamical regime \cite{LIGOScientific:2016aoc,KAGRA:2021vkt,LIGOScientific:2021sio}. Next-generation detectors will go even further and perform high-precision tests of General Relativity (GR), probing a wide variety of questions, from the Kerr hypothesis to deviations from GR and the standard model
\cite{ET:2019dnz,Colpi:2024xhw,Evans:2021gyd}.

Gravitational waveforms have been mostly computed so far in vacuum GR and in some alternative theories of gravity. This will likely prove to be sufficient for detection. However, unbiased parameter estimation as well as accurate tests of GR will likely require understanding possible effects due to matter environments and including them in the analysis pipelines \cite{LISA:2022kgy}. Considerable efforts have been made to quantify the impact of environmental effects on candidate sources, both for baryonic environments like accretion disks and for more exotic dark matter dresses starting from the seminal analysis of Barausse, Cardoso and Pani \cite{Barausse:2014tra}. Astrophysical environments can be thought of as a source of systematic errors in the program of testing GR, but they can be also studied as such through GW. Of particular interest are dark matter-dominated environments, which display different physical signatures depending on the underlying dark sector properties, which could help constrain new fields \cite{Bertone:2018krk}. Newtonian and non-relativistic modelling of the sources' environments have been extensively studied in the literature, to achieve order-of-magnitude estimates and first-order corrections to the GW waveforms. To achieve the level of accuracy required by the science goals of next-generation detectors, relativistic models which take into account the coupling to GR must be developed. Motivated by these considerations, several aspects of relativistic matter effects on gravitational wave observations were  obtained in the last 10 years \cite{Degollado:2014vsa,Brito:2014wla,Macedo:2015ikq,Pani:2015qhr,Macedo:2016wgh,Giudice:2016zpa,Cardoso:2016oxy,Cardoso:2017cfl,
Cardoso:2017cqb,Hannuksela:2018izj,Konoplya:2018yrp,Berti:2019wnn,Cardoso:2019rvt,Bertone:2019irm,Hannuksela:2019vip,Cardoso:2019rou,Kavanagh:2020cfn,Coogan:2021uqv,Cardoso:2021wlq,Pereniguez:2021xcj,Isoyama:2021jjd,Li:2021pxf,Baumann:2021fkf,
Baryakhtar:2022hbu,Speeney:2022ryg,Baumann:2022pkl,Scancella:2022skf,Speri:2022upm,Bamber:2022pbs,DeLuca:2022xlz,Zwick:2022dih,Destounis:2022obl,Singh:2022wvw,Cole:2022yzw,Becker:2022wlo,Kim:2022mdj,Bamber:2023swn,Katagiri:2023yzm,Yang:2023pdu,DeLuca:2023mio,Duque:2023nrf,Camilloni:2023rra,Katagiri:2023umb,Aurrekoetxea:2023jwk,Lenoci:2023gjz,Dai:2023cft,Takahashi:2023flk,Figueiredo:2023gas,Zhao:2023tyo,Traykova:2023qyv,Bhattacharyya:2023kbh,Tomaselli:2023ysb,Kadota:2023wlm,Rahman:2023sof,HerreraMoreno:2023lvm,Nichols:2023ufs,Mukherjee:2023lzn,Zi:2023omh,Ghoshal:2023fhh,Brito:2023pyl,Leong:2023nuk,Destounis:2023ruj,Becar:2023zbl,Shen:2023erj,Duque:2023seg,Speeney:2024mas,Zhang:2024ugv,Santoro:2024tmo,Dyson:2024qrq,Zhang:2024hjr,Montalvo:2024iwq,Alonso-Alvarez:2024gdz,Chowdhury:2024auw,Mitra:2023sny,Karydas:2024fcn,Kavanagh:2024lgq,Zhang:2024hrq,Becker:2024ibd,Yue:2024xhf,Bertone:2024wbn,Bertone:2024rxe,Fischer:2024dte,Zhou:2024vhk,Ireland:2024lye,Zwick:2024xtx,Wilcox:2024sqs,Cannizzaro:2024yee,Jiang:2024lwg,Kumar:2024utz,Cirelli:2024ssz,Rivera:2024xuv,Cannizzaro:2024hdg,Ianniccari:2024ysv,Shen:2024qxv,Spieksma:2024voy,Tan:2024hzw,Duque:2024mfw,Cheng:2024mgl,Boyanov:2024jge,Duque:2024fot,Berti:2024moe,Zwick:2024yzh,Barura:2024uog,DeLuca:2024ufn,DeLuca:2024uju,Cannizzaro:2024fpz,Shadykul:2024ehz,Kamermans:2024ieb,Tahelyani:2024cvk,Yuan:2024duo,Aurrekoetxea:2024cqd,Katagiri:2024wbg,Arana:2024kaz,Katagiri:2024fpn,Mollicone:2024lxy}.

This work will focus on the case of particle dark matter halos, which can grow to large densities in the neighbourhood of a seed BH, generating a characteristic spike profile. This model was derived in the pioneering work of Gondolo and Silk \cite{Gondolo:1999ef} and was generalised to a relativistic formalism in \cite{Sadeghian:2013laa, Ferrer:2017xwm}. The impact of such dark matter distribution on the inspiral phase of a BH binary evolution has been extensively discussed. In \cite{Eda:2014kra}, a first estimate was derived   for the dephasing of the signal due to dynamical friction with respect to vacuum GR, whereas further work refined the computation, pointing out the importance of taking into account the dynamical evolution of the matter distribution \cite{Kavanagh:2020cfn,Coogan:2021uqv,Nichols:2023ufs}. At the same time, dynamical friction effects have been implemented into modular state-of-the-art waveform generators such as \texttt{FastEMRIWaveforms} \cite{Speeney:2022ryg,Speeney:2024mas}. Efforts are underway to achieve a self-consistent, fully relativistic description of GW sources embedded into a dark matter halo, via the development of BH perturbation theory over a non-vacuum background \cite{Cardoso:2021wlq, Cardoso:2022whc}. 
In this article, building upon this program, we will construct a fully relativistic model for an Hernquist matter distribution that surrounds a supermassive black hole, which complete the existing literature \cite{Sadeghian:2013laa, Speeney:2022ryg}. We will develop the theory of perturbations around a spherically symmetric background sourced by anisotropic matter and compare our equations with the literature \cite{Cardoso:2022whc,Liu:2022csl,Duque:2023nrf}. We will then study two typical signatures of gravitational perturbations in the presence of matter: quasi-normal modes and tidal Love numbers. We will compare these quantities computed for a relativistic matter distribution to the values obtained in the literature when considering analytical models inspired by Newtonian physics. 

The quasi-normal spectrum of a dressed BH is altered by the impact of the matter on the gravitational perturbations potential, which yields shifts in the resonant frequencies with respect to vacuum GR \cite{Cardoso:2021wlq}. Black hole spectroscopy has received considerable attention as a key tool to investigate BH properties like the Kerr hypothesis as well as to uncover signatures of beyond-GR physics \cite{Berti:2009kk,Carullo:2021oxn,Cano:2023jbk,Sarkar:2023rhp,Mishra:2023kng}. The detectability prospects of such shifts are currently uncertain \cite{Spieksma:2024voy}. A precise determination of the relativistic effects impact on these quantities is therefore necessary to discuss detectability and possible degeneracies with different theories. Likewise, the static tidal Love numbers of a BH identically vanish in vacuum GR for asymptotically flat spacetimes  \cite{Binnington:2009bb, LeTiec:2020bos, Bhatt:2024yyz, Nair:2024mya, Bhatt:2023zsy}. They are non-zero in the case the compact object is not a BH, for alternative theories to GR or for non-vacuum spacetimes \cite{Cardoso:2019rvt, Cardoso:2019upw, Chakraborty:2023zed}. In this work, we compute the electric-type (odd) and magnetic-type (even) Love numbers for a BH dressed by a relativistic spike. Non-zero tidal Love numbers will contribute to the dephasing of a compact object's inspiral into a BH-spike system, and could be constrained by LISA with high precision \cite{Pani:2019cyc}.

 Through this work we will use geometrized units with $c=G=1$, with the exception of  \ref{sec:dmspikeprofile}, where we will write $G$ explicitly. 

\vspace{6pt} \noindent {\bf Data availability.} We provide three Mathematica notebooks in this \href{https://github.com/gcompere/Tidal-Love-numbers-and-quasi-normal-modes-in-the-Schwarzschild-Hernquist-dark-matter-halo}{Github} repository. The first notebook summarizes the equations of even parity perturbations around a spherically symmetric background sourced by an anisotropic fluid and their properties. The second notebook provides the computation of tidal Love numbers and our stable numerical scheme. The third notebook details the Schwarzschild-Hernquist black hole and provides a solution obeying the dominant energy condition which admits two light rings and one anti-light ring.

\section{Relativistic dark matter spike profiles}
\label{sec:dmspikeprofile}

The first relativistic derivation of a spherically symmetric dark matter (DM) halo around a BH was performed in \cite{Sadeghian:2013laa}, considering a Schwarzschild background. The effects of relativistic rotation were included in \cite{Ferrer:2017xwm} upon upgrading the background to the Kerr metric. In the last 7 years, many subsequent work followed including \cite{Nishikawa:2017chy,Yue:2017iwc,Yue:2018vtk,Hannuksela:2019vip,Cardoso:2019rou,Kavanagh:2020cfn,Coogan:2021uqv,Liu:2022ygf,Cole:2022ucw,Bamber:2022pbs,Speeney:2022ryg,Speeney:2024mas}. In this work, we assume the following: 
\begin{itemize} 
\item The DM halo is composed of cold and collisionless particles modelled by an ideal gas. If instead the DM is interacting, annihilations might deplete the density profile \cite{Vasiliev:2007vh,Shapiro:2016ypb}. If the DM is heated by gravitational interactions, the density could also be depleted in the centre \cite{Ullio:2001fb,Merritt:2003qk,Gnedin:2003rj,Bertone:2005hw}.
\item The BH is formed at the centre of the halo. If it is formed off-centre and then spirals in, the DM spike might not be present and even a central depletion might occur \cite{Ullio:2001fb}.
\item We assume that the BH grows adiabatically. More precisely, we assume that the orbital timescale of DM particles is much faster than the timescale of growth of the central BH. 
\end{itemize}

In the following sections we will start from a non-relativistic spherically symmetric distribution and perform the relaxation to a relativistic profile. We will focus on the Hernquist model. 

\subsection{Spherically symmetric distribution functions}

A collisionless gas of massive particles follow geodesic orbits. Assuming a small backreaction of the DM, the geometry is dictated by the central BH. Assuming spherical symmetry and Einstein gravity, the geometry is the Schwarzschild geometry. Let $u^\mu = \frac{dx^\mu}{d\tau}$ be the four-velocity and $p_\mu = \mu g_{\mu\nu} u^\nu$ the four-momentum of identical particles of mass $\mu$ and proper time $\tau$, $g_{\mu\nu}u^\mu u^\nu=-1$. The free particle Hamiltonian is given by $H(x^\mu,p_\nu)=\frac{1}{2}g^{\mu\nu}(x)p_\mu p_\nu$. One can describe the gas by the phase space distribution function $f=f(x^\mu , p_\nu)$ normalized as $\int f \sqrt{-g} d^4p d^4 x =\mu^{4}$. Liouville theorem states that for a collisionless gas $f$ should be invariant along a geodesic, i.e.
\begin{equation}
	\frac{\drm}{\drm\tau}f(x^\mu(\lambda),p_\nu(\lambda))=\frac{\drm x^\mu}{\drm\lambda}\frac{\partial f}{\partial x^\mu}+\frac{\drm p_\nu}{\drm\lambda}\frac{\partial f}{\partial p_\nu}=\{H,f\}=0.
\end{equation}
This equation is known as the collisionless Boltzmann equation (CBE).  For a given solution, the particle current density can be constructed as
\begin{equation}
	J_\mu= \int f(x,p)p_\mu\sqrt{-g}\,\drm^4p,
	\label{current_def}
\end{equation}
and the mass density can be inferred as $\rho=\sqrt{-J_\mu J^\mu}$. A solution to the CBE can be obtained by finding an adapted phase space coordinate system that makes the equation trivial, so that the solution follows by a choice of boundary conditions. Adapted phase space coordinates are the relativistic angle-action variables $(q^\alpha,J_\alpha)$ \cite{Hinderer2008}. This method has been applied to a Schwarzschild spacetime for the study of accretion phenomena \cite{Rioseco2017,Mach2021}. In the case of spherically symmetric DM halos with a single component mass $\mu$, we assume $f=\mu^{-4}f(E,L)\delta(\mu'-\mu)$ with $f(E,L)$ dimensionless where $E,L$ are the energy per unit mass and the total angular momentum per unit mass. The angular momentum along the $z$-direction per unit mass $L_z$ does not appear in the distribution function by $SO(3)$ symmetry. The CBE is then automatically obeyed since $E=-u_t,L=\sqrt{u_\theta^2+u_\phi^2 \sin^{-2}\theta},L_z=u_\phi$ are constants of geodesic motion. Note that $E,L \geq 0$, $L_z \in \mathbb R$. After performing the coordinate change, integrating over $\mu$ and multiplying by a factor of 4 to take into account both signs of $p_r$, $p_\theta$ for given constants of motion, the particle current density becomes
\begin{equation}
J_\mu = 4\int f(E,L) p_\mu \frac{L}{\sin\theta\sqrt{U R}}dEdLdL_z,
\end{equation}
where $U=L^2-L_z^2 \sin^{-2}\theta$, $R=E^2r^4-\Delta (r^2 +L^2) $, $\Delta=r^2-2 \mbh r$. The components $J_r$, $J_\theta$ are vanishing since bound orbits have a symmetric motion between positive and negative radial or polar velocities. The component $J_\phi$ also vanishes because the integrand is odd under sign reversal of $L_z$. The mass density is given by $\rho=-(1-2\mbh/r)^{-1/2}J_t\mu^{-1}M_{DM}$ where we now normalize the mass with respect to the complete DM mass $M_{DM}$ instead of the individual particle mass. Performing the integral over $L_z$ one obtains
\begin{equation}
\rho(r) = 4\pi M_{DM} (1-\frac{2\mbh}{r})^{-1/2}\int \frac{f(E,L)}{\sqrt{R}} E L dE dL ,
\label{density_formal}
\end{equation}
which agrees with Eq. (3.15) of \cite{Sadeghian:2013laa}. The range of integration of $E,L$ is dictated by the region of existence of bounded orbits that reach $r$, which is known analytically. Such orbits only exist for $r \geq 4GM$ and therefore $\rho(r)=0$ for $r \leq 4GM$. The energy is integrated in the range $E_{\text{min}}(r) \leq E < 1$ with $E_{\text{min}}(r)=(1+2G\mbh/r)/\sqrt{1+6G\mbh/r}$ for $r \geq 6G\mbh$ and $E_{\text{min}}(r)=(1-2G\mbh/r)/\sqrt{1-3G\mbh/r}$ for $4G\mbh \leq r \leq 6G\mbh$. The angular momentum lies in the range $L_{\text{min}}(E) \leq L \leq L_{\text{max}}(E,r)$ with $L_{\text{min}}^2(E)=32(G\mbh)^2/(36E^2-27E^4-8+E(9E^2-8)^{3/2})$ and $L_{\text{max}}(E,r)=r \sqrt{E^2/(1-2G\mbh/r)-1}$. 

The relativistic profile $f(E,L)$ of a DM cloud around a Schwarzschild BH can be fixed by matching via the Eddington inversion method a non-relativistic profile $f'=f'(E'_N,L')$ obtained from N-body simulations without central BH to the relativistic profile with the central Schwarzschild BH \cite{Ferrer:2017xwm}. Here $E'_N$ is the non-relativistic energy related to the relativistic energy as $E'=1+E_N'$. Starting from a non-relativistic distribution $f'(E'_N,L')$ that generates a central Newtonian potential $\Phi(r)$, one can show that the adiabatic growth of a BH conserves the distribution function, i.e. $f(E,L)=f'(E_N',L')$, together with the invariants $I_r$, $I_\phi$ and $I_\theta$, given by
\begin{align}\label{pd}
	I_{r,N}&=\frac{1}{2\pi \mu}\oint p_r\drm r=\frac{1}{2\pi}\oint\sqrt{2E_N'-2\Phi-\frac{L'^2}{r^2}}\,\drm r,\\
	I_{\theta,N}&=\frac{1}{2\pi \mu}\oint p_\theta\drm\theta=\frac{1}{2\pi}\oint\sqrt{L^{'2}-\frac{L_z^{\prime 2}}{\sin^2\theta}}\,\drm\theta=L'-|L'_z|,\\
	I_{\phi,N}&=\frac{1}{2\pi \mu}\oint p_\phi  \drm\phi =\frac{1}{2\pi}\oint L'_z\drm\phi=L'_z,
\end{align}
for a Newtonian system (where the radial integral is defined in between the two roots of the integrand) and by 
\begin{align}
	I_{r,\textrm{Schw}}&=\frac{1}{2\pi \mu}\oint p_r\drm r=\frac{1}{2\pi}\oint\frac{\sqrt{R(r)}}{\Delta(r)}\,\drm r~,\\
	I_{\theta,\textrm{Schw}}&=\frac{1}{2\pi \mu}\oint p_\theta\drm\theta 
	=L-|L_z|~,\\
	I_{\phi,\textrm{Schw}}&=\frac{1}{2\pi \mu}\oint p_\phi\drm\phi=L_z~,
\end{align}
for general relativistic system described by the Schwarzschild metric with $R(r)=E^2r^4-\Delta (r^2 +L^2) $, $\Delta=r^2-2 \mbh r$ (Carter's constant is $Q=L^2-L_z^2$)\footnote{We note that the invariant $I_r^{\rm BH}$ defined in Eq.(4) of \cite{Speeney:2022ryg} is incorrect since it misses a redshift factor. The error dates to the paper \cite{Sadeghian:2013laa} where Eq. (3.19) for the Schwarzschild case is incorrect, while the formula Eq. (3.8) for the Kerr case is correct. This leads to a difference between the derivation of our results and the ones of \cite{Speeney:2022ryg} by a relative factor of $1-2M/6M=2/3$ around the innermost stable circular orbit.}. The angular momenta $L',L'_z$ are automatically matched $L'=L$, $L'_z=L_z$ which is consistent for a spherically symmetric system, while the relativistic energy $E'=E'(E,L)$ is obtained from the matching equation $I_{r,N}(E_N',L)=I_{r,\textrm{Schw}}(E,L)$ once the Newtonian profile $\Phi(r)$ is given.  

\subsection{The Schwarzschild-Hernquist model}\label{sec:SH}

For definiteness, we take as a starting point in this work the Hernquist profile
\begin{equation}
\rho_{\rm H}(r_{\rm s},\rho_0)=\frac{\rho_0}{(r/r_{\rm s})(1+r/r_{\rm s})^{3}}\,,    
\end{equation}
which generates the Newtonian gravitational potential $\Phi_{\rm H}(r)=-(GM_{\rm DM}/r+r_{\rm s})$. Here, $M_{\rm DM}=2\pi \rho_0 r_{\rm s}^{3}$ is the total DM mass in the halo. We can numerically evaluate the relativistic energy $E'=E'(E,L)$ to machine precision. Using Eddington's formula, the dimensionless ergodic distribution function for this model is given by (see e.g. Eq. (4.51) of \cite{Binney-Tremaine}), 
\begin{eqnarray}
f_{\rm H}(\epsilon')=\frac{1}{\sqrt{2}(2\pi)^3 (GM_{\rm DM}r_{\rm s})^{3/2}}\frac{\sqrt{\epsilon'}}{(1-\epsilon')^2}
\left[(1-2\epsilon')(8\epsilon^{\prime 2}-8\epsilon^\prime-3)+\frac{3\sin^{-1}\sqrt{\epsilon'}}{\sqrt{\epsilon'(1-\epsilon')}}\right]\,,
\end{eqnarray}
where $\epsilon'\equiv-(E'-1)r_{\rm s}/GM_{\rm DM}$ is expressed in terms of the relativistic energy $E'(E,L)$. 

\begin{figure}[htp]
\centering
\includegraphics[width=0.49\textwidth]{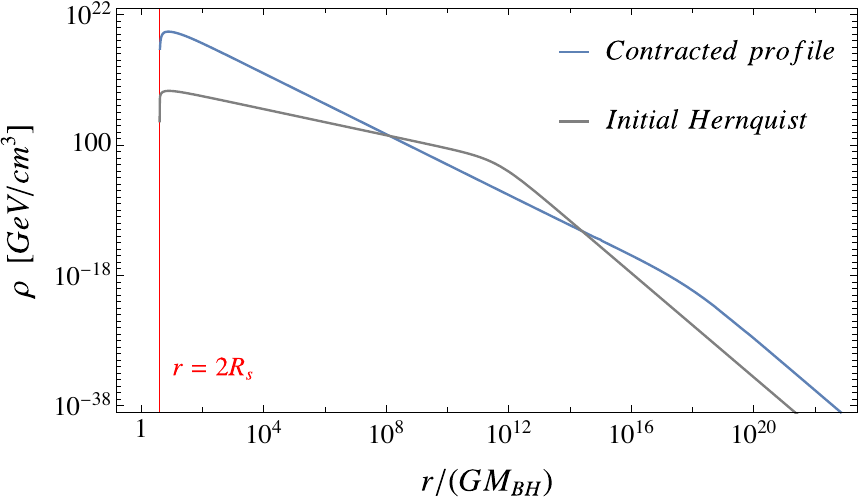}
\caption{DM spike generated by adiabatic contraction from the non-relativistic Hernquist profile. The original profile is shown in gray. The spike has a sharp cutoff at the marginally bound circular orbit, located at radius $2R_{\rm S}=4\mbh$.}
\label{DM_spike_1}
\end{figure}

Following \cite{Sadeghian:2013laa}, we can define the dimensionless quantities $\tilde L=L/\sqrt{r_{\rm s}GM_{\rm DM}}$, $\Tilde \eps=(r_{\rm s}/GM_{\rm DM})(1-E)$ and re-express $\epsilon'=\epsilon'(\Tilde \eps,\tilde L)$. In terms of these quantities, the mass density in Eq. \eqref{density_formal} can be rewritten as 
\begin{align}
\rho=2\pi \frac{G^{3/2}M_{\rm DM}^{5/2}}{r_{\rm s}^{1/2}}\frac{1}{r-2G\mbh}
\int_0^{\Tilde{\eps}_{\rm max}}\drm\Tilde{\eps}\left(1-\frac{GM_{\rm DM}}{r_{\rm s}}\Tilde{\eps}\right)
\int_{\Tilde{L}_{\rm min}^2}^{\Tilde{L}_{\rm max}^2}\drm\Tilde{L}^2\frac{f_{\rm H}({\eps}'(\Tilde{\eps},\Tilde{L})}{\sqrt{\Tilde{L}_{\rm max}^2-{\Tilde{L}^2}}}\,.
\label{density_numerical}
\end{align}
We compute numerically the density, given the above expression, for a reference system with parameters 
\begin{equation}
M_{\rm DM}=10^{12}\,M_{\odot},\,\mbh=10^6\,M_{\odot},\,r_{\rm s}=20\,\textrm{kpc}\,,
\label{Ref1}
\end{equation}
which correspond to a scale density of $\rho_0=0.76\,\textrm{GeV/cm}^3$ and corresponds to plausible astrophysical values \cite{Wilkinson:1999hf}. The result is shown on \ref{DM_spike_1}. The profile obtained from adiabatic contraction admits a sharp spike with cutoff at radius equal to twice the Schwarzschild radius $R_{\rm S}=2G\mbh$ (the location of the marginally bound circular orbit). We also note that the slope of the relativistic profile decreases at a much larger radius $r_{\rm s}'$ than the original $r_{\rm s}$. For the purpose of the subsequent discussion, we have defined the dimensionless variable $x\equiv r/(G\mbh)$. Numerical investigations show that the parameter $x_{\rm s}'$ can be expressed as a combination of other parameters of the problem
\begin{equation}
x_{\rm s}'=\frac{2r_{\rm s} M_{\rm DM}}{G\mbh^{2}}\,,
\label{x_s'}
\end{equation}
which approximates the fitted values of $x_{\rm s}'$ up to $10\,\%$ error. This has been also recently obtained in \cite{Speeney:2024mas}. In the following, we will keep the notation $x_{\rm s}'$ for compactness of the relativistic Schwarzschild-Hernquist profile.
We fit the data via the \texttt{NonLinearModelFit} method of \texttt{Mathematica} to the following model, 
\begin{equation}
\rho=\bar{\rho}(x) \lambda :=\bar{\rho}(x)\left(\frac{M_{DM}}{10^{12}\,M_\odot}\right)^\alpha \left(\frac{\mbh}{10^{6}\,M_\odot}\right)^\beta \left(\frac{r_s}{20\,kpc}\right)^\gamma,
\label{DM_model}
\end{equation}
with the radial dependence of the density profile being,
\begin{equation}
\bar{\rho}(x)=A\left(1-\frac{4}{x}\right)^wx^{-q}\left(1+\frac{x}{x_s'}\right)^{q-4}\,.
\label{DM_model_2}
\end{equation}
The $(\alpha,\,\beta,\,\gamma)$ coefficients are obtained by varying one parameter at a time while keeping the two other fixed. The best-fit values of the $(A,\,w,\,q,\alpha,\,\beta,\,\gamma)$ coefficients are given in \ref{DM_params}. 
It follows that our model approximates the numerical result up to a maximum of $~10\,\%$ error between the fitted model and the numerically obtained data, on nearly the whole range of radii.
\begin{table}[htp]
\def\arraystretch{1.5}
    \centering
    \begin{tabular}{c|c}
                     & Best fit \\
                    \hline
    $A\,\,[\textrm{GeV/cm}^3]$  & $1.85\times10^{22}$ \\
    \hline
  $w$ & $2.45$ \\
    \hline
  $q$ & $2.33$\\
    \hline
    $\alpha$ & $0.33$\\
    \hline
    $\beta$ & $-1.66$ \\
    \hline
    $\gamma$ & $-0.67$ 
    \end{tabular}
    \caption{Best fit parameters, namely $(A,w,q,\alpha,\beta,\gamma)$, for the Schwarzschild-Hernquist BH model, presented in Eq. \eqref{DM_model}.}
    \label{DM_params}
\end{table}

The polynomial ansatz, presented in Eq. \eqref{DM_model} is motivated by the simple model proposed in \cite{Speeney:2022ryg} but contains one important modification. The last term in Eq. \eqref{DM_model_2} is inserted to correctly account for the behaviour of the DM at large radius. Indeed, the profile ought to asymptote to the initial Hernquist power-law profile far away from the region controlled by the adiabatic contraction of the spike, i.e., $\rho \sim r^{-4}$ for $r \gg r_{\rm s}$. This is due to the fact that orbits living at such large distances from the core of the halo will be not perturbed significantly by the adiabatic growth of the central BH. 

The three-parameter model, as advocated in Eq. \eqref{DM_model} with coefficients given in \ref{DM_params} is our proposal for the complete relativistic DM profile obtained from the adiabatic contraction of the Hernquist profile, which we call the Schwarzschild-Hernquist model. It depends upon the total DM mass $M_{\rm DM}$, the BH mass $\mbh$ and the Hernquist shape parameter $r_{\rm s}$. 

We point out that the spike slope we obtain is slightly but significantly steeper than what was concluded in \cite{Speeney:2022ryg}. They obtained $q=1.90$ while we obtain $q=2.33$. We also point out that the $\gamma$ coefficient $\gamma=0.31$ obtained in \cite{Speeney:2022ryg} is very different than the one that we obtain namely $\gamma=-0.67$. A negative value of $\gamma$ is in fact physically motivated: for higher values of $r_s$ and fixed halo mass $M_{DM}$ the DM density gets effectively diluted, with smaller densities for short radii. Thus, an increase in the Hernquist scale radius should yield a smaller density after adiabatic contraction. The discrepancies with the parameters given in \cite{Speeney:2022ryg} come from the fact that the fit performed in that reference is valid only for radii of order $r<\mathcal{O}(10 R_S)$, while it deviates significantly from the contracted profile for larger radii. The analysis carried out in this work requires to consistently model the DM distribution over the whole spacetime. For example, if one was to assume the model given in \cite{Speeney:2022ryg} to be valid up to infinity, the total mass of the spike $M_{DM}=\int_{2R_S}^\infty\drm r\, 4\pi r^2\rho(r)$ would be infinite. Our model compares well with the recent one proposed in \cite{Speeney:2024mas} while this part of our work was finished. The main difference is the fitted value of $\gamma$ which slightly differs.

\begin{figure}[htp]
    \centering
    \includegraphics[width=0.49\textwidth]{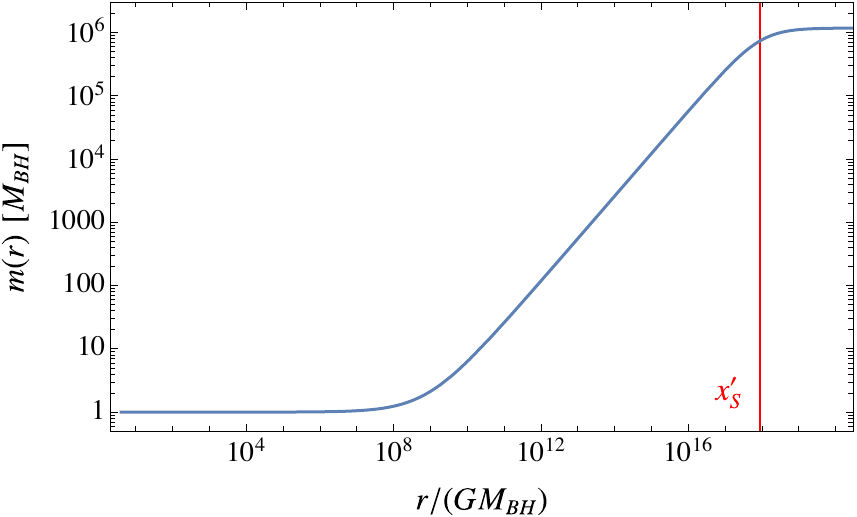}
    \caption{We have plotted the mass distribution in the relativistic DM spike profile in units of the central BH mass for the Schwarzschild-Hernquist model considered above. As evident, for lower radius, the mass profile asymptotes to the mass of the central BH, while at larger radius, it asymptotes to the total mass of the system ($M_{\rm DM}+\mbh$).}
    \label{DM_mass}
\end{figure}

Proceeding further, given the density profile for the DM distribution, the total mass enclosed in a radius $r$ is given by,
\begin{equation}
m(r)=\frac{\pi}{2}R_S^3\lambda\int^r\drm x \, x^2\bar{\rho}(x)\,,
\label{mass_r}
\end{equation}
where we have collected into the $\lambda$ factor all the scaling terms in equation \eqref{DM_model}. We checked that the total mass of the halo differs from the mass of the initial Hernquist profile by only $10\%$ using the fitted parameters given in \ref{DM_params}. Inserting Eq. \eqref{DM_model_2} into Eq. \eqref{mass_r} and using the fact that $x_s' \gg 4$, we obtain the analytic mass density of the Schwarzschild-Hernquist BH model as 
\begin{equation}
m_{\rm SH}(r)=\mbh+\lambda R_S^3 \Tilde{A} \, _2F_1\left(w+1,q+w-2;w+2;-\frac{(x-4) x_s^\prime}{4(x+x_s^\prime)}\right)\Theta(x-4)~, 
\label{mass_SH}
\end{equation}
with, $x=2r/R_S$, $\Theta(x)$ being the Heaviside step function and the factor $\Tilde{A}$ being,
\begin{equation}
\Tilde{A}=\frac{A\,\pi \,x_{\rm s}^{\prime 4}\, 4^{-q-w+2} (x-4)^{w+1} \left(\frac{x+x_{\rm s}^\prime}{x_{\rm s}^\prime}\right)^q
\left(\frac{x_{\rm s}^\prime+4}{x+x^\prime_{\rm s}}\right)^{q+w+1}}{2(w+1) (x_{\rm s}^\prime+4)^4}\,. 
\end{equation}
The above mass profile has been presented in \ref{DM_mass}. Let us remark that the density distribution in Eq. \eqref{DM_model_2} vanishes at $x=4$, the radius of the marginally bound circular orbit. For $2<x<4$, the mass of the spacetime is that of the central BH, and for $x\gg x_{\rm s}'$, it is given by the total mass $(M_{\rm DM}+\mbh)$.  

\subsection{Effective spherically symmetric metric and light-rings}
\label{effective_metric}

We now convert the information about the matter distribution into an effective metric describing the DM surrounding the BH. This can be constructed via an effective averaging of the stress-energy tensor of an ensemble of particles following geodesics on the BH spacetime, given by a matter density $\bar \rho$ \cite{10.2307/1968902,Geralico:2012jt,Cardoso:2021wlq}. Under the simplifying assumption that the eccentricity of such geodesics can be neglected, this can be shown to be equivalent to requiring an anisotropic stress-energy tensor with no radial pressure, 
\begin{equation}
T_\mu^\nu=\mathrm{diag} (-\bar\rho,0,\bar p_t,\bar p_t)\,,
\label{SET}
\end{equation}
which can be coupled to a spherically symmetric metric ansatz via the Einstein equations. Writing,
\begin{equation}
{\overline{ \drm s}}^2=-f(r)\drm t^2+\frac{1}{g(r)}\drm r^2+r^2\drm \Omega^2\,,
\qquad 
g(r) := 1-\frac{2m(r)}{r}\,,
\label{metric_ansatz}
\end{equation}
with $m(r)$ being the mass enclosed in a radius $r$, one gets the following equation relating the metric components
\begin{equation}
\frac{r f'(r)}{f(r)}=\frac{2 m(r)}{r-2m(r)},
\label{Diff_eq_gtt}
\end{equation}
whereas the tangential pressure and the density are related by the 
following relation
\begin{equation}
2 \bar p_t=\frac{m(r)\bar\rho(r)}{r-2m(r)},\qquad \bar \rho(r) = \frac{m'(r)}{4\pi r^2}\,.
\label{PTofRho}
\end{equation}
These equations are consistent with \cite{Cardoso:2022whc}.

\begin{figure}[htp]
\centering
\includegraphics[width=0.49\textwidth]{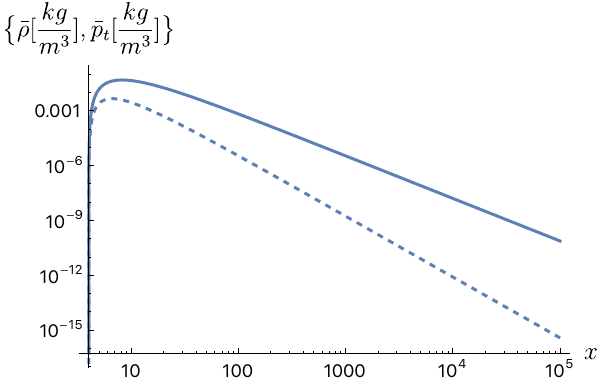}
\includegraphics[width=0.49\textwidth]{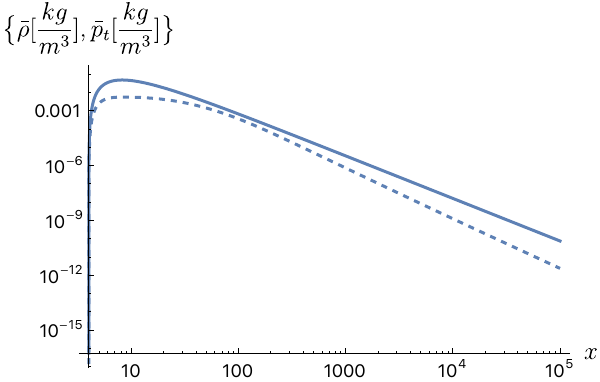}
\caption{Left: Density (solid line) and pressure (dashed line) for the Schwarzschild-Hernquist BH with realistic values \eqref{Ref1}. Right: Density (solid line) and pressure (dashed line) for the Schwarzschild-Hernquist BH with rescaled factor $r_s \mapsto 2.5 \times  10^{-10} r_s $ and other factors unchanged. The latter matter distribution does not violate the dominant energy condition while the metric admits two light rings and one anti-light ring.}
\label{ring}
\end{figure}

A direct consequence of Eq. \eqref{Diff_eq_gtt} is the equality 
\begin{equation}
r-3m(r)=-\frac{\left\{r-2m(r)\right\}r^3}{2f}\dfrac{d}{dr}\left(\frac{f}{r^2}\right)~. 
\end{equation}
Therefore the existence of a light-ring (or anti-light-ring \cite{Cvetic:2016bxi}) in such a geometry, \emph{i.e.}, a locus $r$ such that $\p_r(f/r^2)=0$, is equivalently a locus where 
\begin{align}
r=3m(r)~.
\label{lightring}
\end{align}
Since the matter density vanishes for radii below the standard light-ring $r=3\mbh$, the standard light-ring is present in the geometry. Motivated by recent studies \cite{Guo:2022ghl}, let us study whether a second light-ring might be present. The null, weak and strong energy conditions \cite{hawking_ellis_1973} are obeyed given that $\bar\rho$ and $\bar p_t$ are non-negative in the exterior region $r>2m$. The dominant energy condition reduces to the condition $\bar\rho \geq  \bar p_t$. The constraint only applies when matter is present, for $r \geq 4\mbh$ and it reduces to 
\begin{align}
r \geq \frac{5}{2} m(r)\,.
\label{DEC}
\end{align}
Comparing the existence of a (anti-)light-ring, as in Eq. \eqref{lightring}, and the dominant energy condition constraint \eqref{DEC}, we see that there is a very small window where a light-ring located within matter could exist. However, the dominant energy condition would be only marginally obeyed. This is therefore a very extreme situation which is not physically realistic. We performed a numerical check assuming the  Schwarzschild-Hernquist density \eqref{DM_model}. We obtained that one would need to scale down $r_s$ in the baseline model \eqref{Ref1} as $r_s \mapsto 2.5 \times 10^{-10}r_s$ to achieve such a scenario, which would imply unrealistic DM densities. In that scenario, the pressure $\bar p_{t}$ is enhanced and its value gets closer to the density $\bar \rho$, see \ref{ring}. We therefore conclude that for realistic parameters, there will be one and only one lightring at $r=3\mbh$.

More general matter models are described by an anisotropic fluid with radial pressure, 
\begin{align}
T_{\mu \nu}=\bar \rho\bar  u_{\mu}\bar u_{\nu}+\bar p_{\rm r}\bar v_{\mu}\bar v_{\nu}+\bar p_{\rm t}\bar \Pi_{\mu \nu}~,\label{TmunuAnisobar}
\end{align}
where, $\bar u_{\mu}\bar u^{\mu}=-1$, $\bar v_{\mu}\bar v^{\mu}=+1$, and $\bar v_{\mu}\bar u^{\mu}=0$, with $\Pi_{\mu \nu}=\bar g_{\mu \nu}+\bar u_{\mu}\bar u_{\nu}-\bar v_{\mu}\bar v_{\nu}$. We take $\bar u_\mu = -\sqrt{f(r)}\delta_\mu^t$ and $\bar v^\mu = \sqrt{g(r)}\delta^\mu_r$. Implying, $\bar{u}^{\mu}=\{1/\sqrt{f(r)}\}\delta^{\mu}_{t}$ and $\bar{v}_{\mu}=\{1\sqrt{g(r)}\}\delta^{r}_{\mu}$.

In this case, the Einstein's equations and the conservation of the stress-energy tensor are equivalent to the following three equations
\begin{align}
\bar \rho(r) &= \frac{m'(r)}{4\pi r^2}~, 
\qquad 
\frac{f'(r)}{f(r)}=-\frac{2 (m(r)+4\pi \bar p_{\rm r}(r) r^3)}{r(2 m(r)-r)}~, 
\\ 
\bar p_{\rm t}(r) &= \frac{-2 r \bar p_{\rm r}(r)(1+2\pi r^2 (\bar p_{\rm r}(r)+\bar \rho(r)))-r^2 \bar p_{\rm r}'(r)+m(r) (3 \bar p_{\rm r}(r)-\bar \rho(r)+2 r \bar p_{\rm r}'(r))}{4 m(r)-2 r}. 
\end{align}
The generic solution is given in terms of the (arbitrary) radial mass aspect $m(r)$ and the radial pressure $\bar p_{\rm r}(r)$. These equations disagree with Ref. \cite{Duque:2023nrf} (see equation (9.6) of this paper). In the case $\bar p_{\rm r}=0$, we recover Eqs. \eqref{Diff_eq_gtt}-\eqref{PTofRho}. 

\section{Perturbed Einstein's equations in the presence of matter}

Having discussed the background geometry as well as the matter content, describing a BH immersed in a non-relativistic/relativistic DM distribution, in this section we discuss the perturbation of the same. This involves perturbations of the geometry, described by $\delta g_{\mu \nu}$, as well as perturbation of the matter, captured by $\delta T_{\mu \nu}$. The corresponding equations relating the perturbed metric to the perturbed matter sector, are the linearized Einstein's equations:
\begin{align}\label{pertEin}
\delta G_{\mu \nu}=8\pi G \delta T_{\mu \nu}~,
\end{align}
and the perturbed conservation equation for the matter stress-energy tensor, i.e., 
\begin{align}\label{pertmatter}
\delta \left(\nabla_{\mu}T^{\mu \nu}\right)=0~.
\end{align}
In what follows we will present the perturbed matter stress-energy tensor, as well as the perturbed Einstein tensor, in terms of certain fundamental observables associated with the matter and the metric. We start with the perturbed matter stress-energy tensor and shall decompose it into axial and polar parts. 
\subsection{Perturbation of the matter stress-energy tensor}

Given the fact that the background matter distribution was anisotropic ($p_{r}=0\neq p_{t}$), it only makes sense to write down 
the perturbed stress-energy tensor of an anisotropic fluid, which takes the following form 
\begin{align}
T_{\mu \nu}=\rho u_{\mu} u_{\nu}+ p_{\rm r} v_{\mu} v_{\nu}+ p_{\rm t} \Pi_{\mu \nu}~,\label{TmunuAniso}
\end{align}
where $ u_{\mu} u^{\mu}=-1$, $v_{\mu}v^{\mu}=+1$, $ v_{\mu} u^{\mu}=0$ and $\Pi_{\mu \nu}= g_{\mu \nu}+ u_{\mu} u_{\nu}- v_{\mu} v_{\nu}$. Here, the metric is perturbed as $g_{\mu\nu}=\bar g_{\mu\nu}+h_{\mu\nu}$ and the scalars as $\rho=\bar \rho+\delta \rho$, $p_{\rm t}=\bar p_{\rm t}+\delta p_{\rm t}$, $p_{\rm r}=\bar p_{\rm r}+\delta p_{\rm r}$. Note that, the above anisotropic stress-energy tensor can describe the DM halo, provided we set $\bar{p}_{r}=0$. The perturbations of the fluid four-velocity $u^{\mu}$ and the auxiliary four-vector $v^{\mu}$ are defined as variations of the contravariant components of the respective fluid vectors, 
\begin{equation}
u^\mu = \bar u^\mu + \delta u^\mu , \qquad v^\mu = \bar v^\mu + \delta v^\mu\,. 
\end{equation}
The perturbations of the covariant components of the fluid vectors are instead given by 
\begin{align}
\delta (u_\mu)=h_{\mu\nu}u^\nu + \bar g_{\mu\nu}\delta u^\nu,\qquad \delta (v_\mu)=h_{\mu\nu}v^\nu + \bar g_{\mu\nu}\delta v^\nu\,. 
\end{align}
Due to the metric perturbation, it is in general not equivalent to impose either $\delta u^\mu =0$ or $\delta (u_\mu)=0$. We will refer to the first choice as the 'up' choice and refer to the second choice as the 'down' choice for cancelling the variation of the fluid vector. It corresponds to physically different dynamics for the fluid.

\subsection{Perturbation equations: odd sector}
\label{perturbation_odd}

The odd metric perturbations in Regge-Wheeler gauge are defined in the standard fashion as 
\begin{align}
h_{t\theta}=h_{\theta t}=-\frac{h_0(t,r)}{\sin\theta}\partial_\phi Y_{\ell m},
\qquad 
h_{t\phi}=h_{\phi t}=h_0(t,r)\sin\theta\partial_\theta Y_{\ell m}, 
\\
h_{r\theta}=h_{\theta r}=-\frac{h_1(t,r)}{\sin\theta}\partial_\phi Y_{\ell m},
\qquad 
h_{r\phi}=h_{\phi r}=h_1(t,r)\sin\theta\partial_\theta Y_{\ell m}. 
\end{align}
Besides the metric perturbations, as presented above, there will also be perturbations associated with the fluid elements, in particular the fluid four-velocity $u^{\mu}$ and auxiliary vector field $v^{\mu}$. In the axial sector, perturbations of each of them will introduce one extra degree of freedom. However, it turns out that these perturbation variables have different values for $\delta u^{\mu}=0=\delta v^{\mu}$, compared to $\delta u_{\mu}=0=\delta v_{\mu}$. As a consequence, the master equations describing the QNMs as well as the static TLNs are different, depending on whether the perturbation of the covariant or, the contravariant components of $u^{\mu}$ and $v^{\mu}$ have been set to zero. Following which we discuss these two cases separately and describe them as the `down' and the `up' definitions, respectively. 

\subsubsection{The `up' master equations}

The only odd perturbations in the matter sector are the odd perturbations of $\delta u^\mu$ and $\delta v^\mu$, which are given by 
\begin{align}
\delta u^\theta &=-\frac{\sqrt{f(r)}}{4\pi(\bar p_{\rm t}+\bar\rho)r^2 }\frac{U^{\rm{up}}(r)e^{-i\omega t}}{\sin\theta}\partial_\phi Y_{\ell m}\,, 
\qquad 
\delta u^\phi=\frac{\sqrt{f(r)}}{4\pi(\bar p_{\rm t}+\bar\rho) r^2\sin^2\theta }U^{\rm{up}}(r)e^{-i\omega t}\sin\theta\partial_\theta Y_{\ell m}\,,
\\
\delta v^\theta &=-\frac{\sqrt{g(r)}}{4\pi(\bar p_{\rm t}+\bar\rho)r^2 }\frac{V^{\rm{up}}(r)e^{-i\omega t}}{\sin\theta}\partial_\phi Y_{\ell m}\,, 
\qquad 
\delta v^\phi=\frac{\sqrt{g(r)}}{4\pi(\bar p_{\rm t}+\bar \rho)r^2\sin^2\theta }V^{\rm{up}}(r)e^{-i \omega t}\sin\theta \partial_\theta Y_{\ell m}\,.
\end{align}
Therefore, the perturbed non-zero components of the matter stress-energy tensor associated with the axial perturbation become,
\begin{align}
\delta T_{t\phi}^{\rm axial}&=e^{-i\omega t}\left[-\frac{fU^{\rm up}}{4\pi}-\bar{\rho}h_{0}(r)\right]\sin \theta \partial_{\theta}Y_{\ell m}~,
\\
\delta T_{t\theta}^{\rm axial}&=-e^{-i\omega t}\left[-\frac{fU^{\rm up}}{4\pi}-\bar{\rho}h_{0}(r)\right]\frac{\partial_{\phi}Y_{\ell m}}{\sin \theta}~,
\\
\delta T_{r\phi}^{\rm axial}&=e^{-i\omega t}\left[\bar{p}_{r}h_{1}(r)+\frac{(\bar{p}_{r}-\bar{p}_{t})V^{\rm up}}{4\pi(\bar{\rho}+\bar{p}_{t})}\right]\sin \theta \partial_{\theta}Y_{\ell m}~,
\\
\delta T_{r\theta}^{\rm axial}&=-e^{-i\omega t}\left[\bar{p}_{r}h_{1}(r)+\frac{(\bar{p}_{r}-\bar{p}_{t})V^{\rm up}}{4\pi(\bar{\rho}+\bar{p}_{t})}\right]\frac{\partial_{\phi}Y_{\ell m}}{\sin \theta}~.
\end{align}
One can verify that the above components of the perturbed matter stress-energy tensor identically satisfies Eq. \eqref{pertmatter}.

Substituting these components to the right hand side of Eq. \eqref{pertEin}, and using the perturbed Einstein tensor from the metric perturbation described above, 
we obtain three independent differential equations, these correspond to the $(\theta,\phi)$, $(r,\phi)$ and $(t,\phi)$ components of Eq. \eqref{pertEin}, respectively. The $(\theta,\phi)$ component of the perturbed Einstein's equations is homogeneous and relates $h_0$ with $h_1$, 
\begin{align}\label{eq1}
-i\omega h_{0}-\sqrt{fg}\dfrac{d}{dr}\left(\sqrt{fg}h_{1}\right)=0~.
\end{align}
After using the above equation and the tortoise coordinate $dr_{*}\equiv dr/\sqrt{f g}$, the $(r,\phi)$ component of the perturbed Einstein's equation can be written as the Regge-Wheeler equation with source for the function $\Psi_{\rm RW}(r_*)=(\sqrt{fg}/r)h_{1}$, 
\begin{align}\label{eq2}
\dfrac{d^{2}{\Psi}_{\rm RW}}{dr_{*}^{2}}+\left[\omega^{2}-V_{\rm RW}^{\rm up}\right]{\Psi}_{\rm RW}
=\frac{4f\sqrt{fg}}{r}\left(\frac{\bar p_{\rm t}-\bar p_{\rm r}}{\bar \rho+\bar p_{\rm t}}\right)V^{\text{up}}~, 
\end{align}
where, 
\begin{align}
V_{\rm RW}^{\rm up}(r)&=f\left[\frac{\ell (\ell+1)}{r^2}-\frac{6m}{r^3}+4\pi(\bar\rho+4\bar p_{\rm t}-5\bar p_{\rm r}) \right]
\nonumber
\\
&=f\left[\frac{\ell(\ell+1)}{r^{2}}-\frac{3}{2r}\left(g'+g\frac{f'}{f}\right)+8\pi\left(-\bar\rho-\bar p_{\rm r}+2\bar p_{\rm t}\right)\right]~.
\label{RW_Pot_Up}
\end{align}
The third and the final perturbed Einstein's equation comes from the $(t,\phi)$ component of Eq. \eqref{pertEin}, which can be written as,
\begin{equation}\label{eq3}
g\left[h_0''+i\omega \left(h_1'+\frac{2}{r}h_1\right)\right]-4\pi r\left(\bar p_{\rm r}+\bar \rho\right)\left(h_0'+i\omega h_1\right)-\left[\frac{(\ell+2)(\ell-1)}{r^2}+\frac{2g}{r^2}+8\pi \left(\bar\rho -\bar p_{\rm r} +2\bar p_{\rm t}\right)\right]h_0 =4 f U^{\rm{up}}\,.
\end{equation}
In summary, the three non-trivial field equations in the context of axial perturbations are given by the Eqs. \eqref{eq1}, \eqref{eq2} and \eqref{eq3}. In the absence of perturbations, in the sense that the contravariant components of the perturbed fluid velocity vanish, i.e., $\delta u^\mu =\delta v^\mu = 0$, we set 
\begin{align}
U^{\rm{up}}=0=V^{\rm{up}}\,. 
\end{align} 
This sets to zero the source terms in Eqs. \eqref{eq2}-\eqref{eq3}. 

\subsubsection{The `down' master equations}

The absence of perturbations, in the sense that the covariant components of the perturbed fluid velocity vanish, i.e., $\delta (u_\mu)=\delta (v_\mu)=0$, gives instead 
\begin{align}\label{deltaudown}
U^{\rm{up}}=-\frac{4\pi(\bar\rho +\bar p_{\rm t})}{f}h_0~,
\qquad 
V^{\rm{up}}=-4\pi(\bar \rho+\bar p_{\rm t})h_{1}~.
\end{align}
In that case, it is natural to rewrite Eqs. \eqref{eq2}-\eqref{eq3} instead as 
\begin{align}\label{eq2bis}
\dfrac{d^{2}{\Psi}_{\rm RW}}{dr_{*}^{2}}+\left[\omega^{2}- V_{\rm RW}^{\rm{down}}\right]{\Psi}_{\rm RW}=0~,
\end{align}
where, we have the modified potential, 
\begin{equation}
 V_{\rm RW}^{\rm{down}}(r)=f\left[\frac{\ell(\ell+1)}{r^2}-\frac{6m}{r^3}+4\pi(\bar \rho-\bar p_{\rm r})\right]~,
\label{RW_Pot_Down}
\end{equation}
and the following differential equation for $h_{0}$
\begin{equation}\label{eq3bis}
g \left[h_0''+i\omega \left(h_1'+\frac{2}{r}h_1\right)\right]-4\pi r\left(\bar p_{\rm r}+\bar \rho\right)\left(h_0'+i\omega h_1\right)-\left[\frac{(\ell+2)(\ell-1)}{r^2}+\frac{2g}{r^2}-8\pi\left(\bar \rho+\bar p_{\rm r}\right)\right]h_0 =0~.
\end{equation}
Therefore, after imposing Eq. \eqref{deltaudown}, we obtain three homogeneous field equations, given by Eqs. \eqref{eq1}, \eqref{eq2bis}, and \eqref{eq3bis}.

We note that for a background isotropic fluid distribution, with $\bar p_{\rm r} = \bar p_{\rm t}$, we have $V_{\rm RW}^{\rm up}(r)=V_{\rm RW}^{\rm down}(r)$. Thus, only for a background anisotropic fluid, the potentials of the Regge-Wheeler equation will differ between the `up' and the `down' definitons. This implies that the QNM spectrum will also differ, depending on whether one fixes $\delta v^\mu=0=\delta u^{\mu}$ or $\delta v_\mu =0=\delta u_{\mu}$. We now move on to determining the perturbation equations in the even sector.

\subsection{Perturbation equations: even sector}
\label{sec:even}

The even metric perturbations in the Regge-Wheeler gauge are defined in the standard fashion as 
\begin{align}
h_{tt}&= e^{-i\omega t} f(r) H_0(r)Y_{\ell m}~, 
\qquad
h_{tr}= h_{rt} = e^{-i\omega t} H_1(r)Y_{\ell m}~, 
\qquad
h_{rr}= e^{-i\omega t} \frac{H_2(r)}{g(r)}Y_{\ell m}~, \\
h_{\theta\theta}&= e^{-i\omega t} r^2 K(r)Y_{\ell m}~,
\qquad 
h_{\phi\phi}= e^{-i\omega t} r^2 \sin^2\theta K(r)Y_{\ell m}~. 
\end{align}
The perturbed stress-energy tensor associated with the matter distribution has already been defined in Eq. \eqref{TmunuAniso}. In the even sector, all the scalar perturbations will contribution, which are described with the following variables: 
\begin{align}
\delta \rho(t,r,\theta,\phi) &= e^{-i\omega t}\delta \rho(r)Y_{\ell m}~, \\
\delta p_{\rm t}(t,r,\theta,\phi) &= e^{-i\omega t}\delta p_{\rm t}(r)Y_{\ell m}~, \\
\delta p_{\rm r}(t,r,\theta,\phi) &= e^{-i\omega t}\delta p_{\rm r}(r)Y_{\ell m}~.
\end{align}
In order to close our system of equations, and to avoid unnecessary complications, we shall not consider arbitrary variations of the vectors $u^\mu$, $v^\mu$. Further, as we will demonstrate, unlike the case of axial perturbations, here there is no difference between the `up' and the `down' components for the polar sector. 
 
We start by considering perturbations of the contravariant components of the fluid four-vector $u^{\mu}$ and the auxiliary four-vector $v^{\mu}$, such that, the non-zero components of $\delta u^{\mu}$ and $\delta v^{\mu}$ are,
\begin{align}
\delta u^{t}&=\frac{H_{0}(r)}{2\sqrt{f}}Y_{\ell m}e^{-i \omega t}~,
\quad
\delta u^{r}=\frac{g}{\sqrt{f} r^2}\frac{D^{\rm up}(r)Y_{\ell m}}{16\pi(\bar{\rho}+\bar{p}_{r})}e^{-i \omega t}~,
\quad
\delta u^{\theta}=\frac{\sqrt{f}C^{\rm up}(r)\partial_{\theta}Y_{\ell m}}{4\pi r^{2}(\bar\rho+\bar p_{t})}e^{-i \omega t}~,
\nonumber
\\
\delta u^{\phi} &=\frac{\sqrt{f}C^{\rm up}(r)\partial_{\phi}Y_{\ell m}}{4\pi r^{2}\sin^{2}\theta(\bar\rho+\bar p_{t})}e^{-i \omega t}~,
\\
\delta v^{t}&=\frac{\sqrt{g}}{f}\left[\frac{D^{\rm up}}{16\pi r^2 (\bar{\rho}+\bar{p}_{r})}+H_{1}\right]Y_{\ell m}e^{-i \omega t}~,
\quad
\delta v^{r}=-\frac{\sqrt{g}H_{2}}{2}Y_{\ell m}e^{-i \omega t}~,
\quad
\delta v^{\theta}=0=\delta v^{\phi}~.
\end{align}
Note that, given the spherical symmetry of the background geometry, it follows that the perturbed vectors $\delta u^{\mu}$ and $\delta v^{\mu}$ are each characterized by three independent even harmonics. Among these six variables, three are fixed in terms of metric perturbations, thanks to the following three constraints: $\delta (u^{\mu}u_{\mu})=0$, $\delta (u^{\mu}v_{\mu})=0$, and $\delta (v_{\mu}v^{\mu})=0$. Among the other three variables, we have $D^{\rm up}$ and $C^{\rm up}$, while the sixth one, associated with the $\delta v^{\theta}$ and $\delta v^{\phi}$ components, has been set to zero for simplicity. Note that unlike the axial sector, here the components of $\delta u^{\mu}$ and $\delta v^{\mu}$ are related to metric perturbations and hence cannot be set to zero. 

The 'down' components, i.e., the components of $\delta u_\mu$ and $\delta v_\mu$, on the other hand, reads, 
\begin{align}
\delta u_{t}&=\frac{\sqrt{f}H_{0}(r)}{2}Y_{\ell m}~,
\quad
\delta u_{r}=\frac{H_{1}}{\sqrt{f}}+\frac{D^{\rm up}(r)Y_{\ell m}}{16\pi r^{2} \sqrt{f}(\bar{\rho}+\bar{p}_{r})}~,
\quad
\delta u_{\theta}=e^{-i\omega t}\frac{\sqrt{f}C^{\rm up}}{4\pi (\bar{\rho}+\bar{p}_{t})}\partial_{\theta}Y_{\ell m}~,
\nonumber
\\
\delta u_{\phi}&=e^{-i\omega t}\frac{\sqrt{f}C^{\rm up}}{4\pi (\bar{\rho}+\bar{p}_{t})}\partial_{\phi}Y_{\ell m}~,
\\
\delta v_{t}&=-\frac{\sqrt{g}D^{\rm up}}{16\pi r^{2}(\bar{\rho}+\bar{p}_{r})}Y_{\ell m}~,
\qquad
\delta v_{r}=\frac{H_{2}}{2\sqrt{g}}Y_{\ell m}~,
\qquad
\delta v_{\theta}=0=\delta v_{\phi}~.
\end{align}
Thus, if we set $\delta u^{\mu}=0=\delta v^{\mu}$, it follows that $\delta u_{\mu}=0=\delta v_{\mu}$. Hence there is no distinction between `up' and the `down' master equations for the polar sector. 

Given the metric perturbations, as well as fluid perturbations, the non-zero components of the perturbed matter stress-energy tensor in the polar sector become,
\begin{align}
\delta T_{tt}^{\rm polar}&=e^{-i\omega t}\left[f\delta \rho-fH_{0}\bar{\rho}\right]Y_{\ell m}~,
\\
\delta T_{tr}^{\rm polar}&=-e^{-i\omega t}\left[\frac{D^{\rm up}}{16\pi r^2}+\bar{\rho}H_{1}\right]Y_{\ell m}~,
\\
\delta T_{rr}^{\rm polar}&=e^{-i\omega t}\left[\frac{\delta p_{r}}{g}+\frac{\bar{p}_{r}H_{2}}{g}\right]Y_{\ell m}~,
\\
\delta T_{t\theta}^{\rm polar}&=e^{-i\omega t}\left[-\frac{fC^{\rm up}(r)}{4\pi}\right]\partial_{\theta}Y_{\ell m}~,
\\
\delta T_{t\phi}^{\rm polar}&=e^{-i\omega t}\left[-\frac{fC^{\rm up}(r)}{4\pi}\right]\partial_{\phi}Y_{\ell m}~,
\\
\delta T_{\theta \theta}^{\rm polar}&=e^{-i\omega t}\left[r^{2}\delta p_{t}+r^{2}\bar{p}_{t}K(r)\right]Y_{\ell m}
=\textrm{cosec}^{2}\theta\,\delta T_{\phi \phi}^{\rm polar}~.
\end{align}
One can verify that the above expressions exactly match with those reported in \cite{Cardoso:2022whc}. 

Having presented the non-zero perturbed components of the matter stress-energy tensor, we now report the perturbed Einstein's equations associated with the polar perturbation. We summarize these equations below. 
\begin{itemize}

\item We start with the $(\theta,\phi)$  component of the perturbed Einstein's equations which directly yields 
\begin{align}
H_{0}(r)=H_{2}(r)\equiv H(r)~,
\label{eqH}
\end{align}
after factoring out the overall factors depending on $e^{-i \omega t}$ and spherical harmonics. The subtraction of the $(\theta,\theta)$ and $\csc^2\theta$ times the $(\phi,\phi)$ components of the perturbed Einstein's equations yields the same equation. 
We will assume that Eq. \eqref{eqH} is obeyed in order to simplify the remaining perturbed Einstein's equations in what follows. 

\item The $(t,t)$ component of the perturbed Einstein's equations read (with $H_{2}$ replaced by $H$),
\begin{align}\label{pertGt}
K''+\left(\frac{3}{r}+\frac{g'}{2g}\right)K'
-\frac{(\ell-1)(\ell+2)}{2r^{2}g}K-\frac{H'}{r}-\left(\frac{\ell(\ell+1)+2-16\pi r^{2}\bar{\rho}}{2r^{2}g}\right)H=-\frac{8\pi \delta \rho}{g}~.
\end{align}

\item The $(t,r)$ component of the perturbed Einstein's equations become (with $H_{2}$ replaced by $H$), 
\begin{align}
i\omega K'-\frac{i\omega}{r}H+\frac{\ell(\ell+1)}{2r^{2}}H_{1}+\frac{i\omega}{2r g}\left(-8\pi r^{2}\bar{p}_{r}+3g-1\right)K=-\frac{1}{2r^2}D^{\rm up}~.
\end{align}

\item The $(r,r)$ component of the perturbed Einstein's equations yield (with $H_{2}=H_{0}=H$), 
\begin{align}
g H'+\frac{2i\omega g}{f}H_{1}-\frac{\omega^2 r}{f}K+\frac{(\ell-1)(\ell+2)}{2r}(K-H)+8\pi r\bar p_r H-\frac{8\pi r^2\bar{p}_{r}+1+g}{2}K'=-8\pi r \delta p_{r}~.
\end{align}

\item The $(\theta,\theta)$ + $(\phi,\phi)$ component of the perturbed Einstein's equations yield (with $H_{2}=H_{0}=H$),
\begin{align}\label{pertGtheta}
g\left(H''-K''\right)&+\frac{2i\omega g}{f}H_{1}'
+\frac{H'}{r}\left\{-4\pi r^{2}\left(\bar{\rho}-3\bar{p}_{r}\right)+2\right\}
+\frac{K'}{r}\left\{4\pi r^{2}\left(\bar{\rho}-\bar{p}_{r}\right)-\left(1+g\right)\right\}
\nonumber
\\
&-\frac{\omega^{2}}{f}\left(K+H\right)+\frac{i\omega}{rf}\left\{-8\pi r^{2}\bar{\rho}+1+g\right\}H_{1}
+16\pi \bar p_t H=-16\pi \delta p_t~.
\end{align}

\item The $(t,\theta)$ component yields, 
\begin{align}
gH_{1}'+i\omega\left(K+H\right)
+\left\{1-g-4\pi r^{2}\left(\bar{\rho}-\bar{p}_{r}\right) \right\}\frac{H_{1}}{r}=-4fC^{\rm up}~.
\end{align}

\item The $(r,\theta)$ component yields,  
\begin{align}
H'-K'+\frac{i\omega }{f}H_{1}+\frac{1-g+8\pi r^{2}\bar{p}_{r}}{rg}H=0~.
\end{align}

\item The $(t,\phi)$ and $(r,\phi)$ components will yield identical equations. Hence no new equations can be obtained. 

\end{itemize}

Besides the perturbed Einstein's equations, we also require equations for the matter sector, which are obtained by considering the components of Eq. \eqref{pertmatter}, yielding:
\begin{itemize}

\item The $t$ component of the perturbed conservation equation reads
\begin{align}\label{conservpertt}
\delta \rho&=\left(\frac{g}{16\pi i\omega r^{2}}\right){D^{\rm up}}^{'}+\left(\frac{gf'+fg'}{32\pi i\omega r^{2}f}\right)D^{\rm up}-\left(\frac{\ell(\ell+1)f}{4\pi i\omega r^{2}}\right)C^{\rm up}-\frac{\bar \rho+\bar p_r}{2} H
\nonumber
\\
&-\left\{ \frac{(\bar \rho+\bar p_r)(1+3 g +8\pi r^2 \bar p_r)}{4g} + \frac{r \bar p_r'}{2} \right\}K\,.
\end{align}

\item The $r$ component of the perturbed conservation equation reads
\begin{align}\label{conservpertr}
\delta p_{r}'&=\left(\frac{i\omega}{16\pi r^{2}f}\right)D^{\rm up}-\left(\frac{4f+rf'}{2rf}\right)\delta p_{r}+\left(\frac{2}{r}\right)\delta p_{t}-\left(\frac{f'}{2f}\right)\delta \rho+\left(\frac{i\omega (\bar{p}_{r}+\bar{\rho})}{f}\right)H_{1}+\left(\frac{\bar{p}_{r}+\bar{\rho}}{2}\right)H'
\nonumber
\\
&+\left[\frac{r\bar{p}_{r}'}{2}+\frac{(\bar{\rho}+\bar{p}_{r})r f'}{4f}\right]K'\,.
\end{align}

\item The $\theta$ and the $\phi$ component of the perturbed conservation equation yields identical results, which read
\begin{align}\label{conservperttheta}
\delta p_t - \frac{i \omega}{4\pi}C^{\rm up} -\frac{\bar \rho +\bar p_r}{2}H=0~. 
\end{align}

\end{itemize}

Thus in total we have \emph{nine} equations, \emph{six} from the gravitational sector and \emph{three} from the matter sector. We also have \emph{eight} unknowns to fix --- (a) the gravitational perturbation yields \emph{three} unknowns: $K$, $H_{1}$ and $H$, (b) the perturbation of the matter sector has \emph{five} unknowns: $\delta \rho$, $\delta p_{r}$, $\delta p_{t}$, $D^{\rm up}$ and $C^{\rm up}$. Thus it might appear that the system if equations over determine the system of unknowns, however, as we will demonstrate that is not the case.

\subsubsection{Master system of equations for polar perturbations}

Let us now determine the master system of equations in the context of polar gravitational perturbation. First of all, in the present situation, we already have $H_{0}=H_{2}\equiv H(r)$, and the following four first order differential equations 
\begin{align}
K'&-\frac{H}{r}+\frac{\ell(\ell+1)}{2i\omega r^{2}}H_{1}+\left(\frac{-8\pi r^{2}\bar{p}_{r}+3g-1}{2r g}\right)K=-\frac{1}{2i\omega r^2}D^{\rm up}~,
\label{zerillimat01}
\\
H'&+\frac{2i\omega}{f}H_{1}-\frac{\omega^2 r}{fg}K+\frac{(\ell-1)(\ell+2)}{2rg}(K-H)+\frac{8\pi r\bar p_r}{g} H-\frac{8\pi r^2\bar{p}_{r}+1+g}{2g}K'=-\frac{8\pi r \delta p_{r}}{g}~,
\label{zerillimat02}
\\
H_{1}'&+\frac{i\omega}{g}\left(K+H\right)
+\left\{1-g-4\pi r^{2}\left(\bar{\rho}-\bar{p}_{r}\right) \right\}\frac{H_{1}}{rg}=-\frac{4f}{g}C^{\rm up}~,
\label{zerillimat03}
\\
H'&-K'+\frac{i\omega }{f}H_{1}+\frac{1-g+8\pi r^{2}\bar{p}_{r}}{rg}H=0~.
\label{zerillimat04}
\end{align}
In addition, there are two second order differential equations, that we will analyse later on. Substitution of $K'$ from Eq. \eqref{zerillimat01} to Eq. \eqref{zerillimat04} yields the following equation,
\begin{align}
H'+\left(\frac{-8\pi r^{2}\bar{p}_{r}+3g-1}{2r g}\right)K+\left[\frac{\ell(\ell+1)}{2i\omega r^{2}}+\frac{i\omega}{f}\right]H_{1}+\left(\frac{1-2g+8\pi r^{2}\bar{p}_{r}}{rg}\right)H=-\frac{D^{\rm up}}{2i\omega r^2}~.
\label{zerillimat05}
\end{align}
Substitution of the above expression for $H'$ in Eq. \eqref{zerillimat02}, yields the following algebraic relation between these three gravitational perturbation quantities,
\begin{align}\label{zerillimat06}
\Big[\ell(\ell+1)+16 \pi r^2\bar p_r &-\frac{2\omega^{2}r^{2}}{f}-\frac{3g}{2} -\frac{\left(1+8\pi r^{2}\bar{p}_{r}\right)^2}{{2g}} \Big]K+\left[\frac{2i\omega rg}{f}+\frac{\ell(\ell+1)(8\pi r^{2}\bar{p}_{r}+1-g)}{2i\omega r} \right]H_{1}
\nonumber
\\
&-\left[(\ell+2)(\ell-1)+8\pi r^{2}\bar{p}_{r}+3(1-g)\right]H=-16\pi r^{2}\delta p_{r}+\frac{D^{\rm up}}{2i\omega r}\left(g-1-8\pi r^{2}\bar{p}_{r}\right)~.
\end{align}
This implies that the perturbation variable $H$ can be expressed in terms of $(K,H_{1},\delta p_{r},D^{\rm up})$ and hence we are left with three independent first order equations. However, we can now substitute the algebraic solution $H=H(K,H_{1},\delta p_{r},D^{\rm up})$ into Eq. \eqref{zerillimat02}, which can be shown to be equivalent with the equation obtained after using the other two first order equations as well as the matter equations, namely Eqs. \eqref{conservpertt} and \eqref{conservpertr}. Thus, one of the first order equation, in particular Eq. \eqref{zerillimat02}, is therefore redundant.  

The remaining two second order perturbation equations in the gravitational sector, as presented in Eq.\eqref{pertGt} and Eq.\eqref{pertGtheta} can be shown to be compatible with the first order equations, provided two additional constraints are satisfied. After analysis, these two additional constraints are found to be nothing else than the matter equations, Eqs. \eqref{conservpertt} and \eqref{conservperttheta}, respectively. The second order equations are therefore also redundant and hence can be discarded. 

In summary, we are left with two algebraic equations: one from the matter sector, Eq. \eqref{conservperttheta}, and one from the Einstein sector, Eq. \eqref{zerillimat06}, as well as four first order equations: two from the matter sector, Eqs. \eqref{conservpertt} and \eqref{conservpertr}, and two from the Einstein sector, let us say Eqs. \eqref{zerillimat01} and \eqref{zerillimat03}. We choose to solve these equations for the variables $K$, $H_1$, $\delta \rho$ and $D^{\rm up}$ and to algebraically solve the two constraints for $H$ and $C^{\rm up}$ (which is valid for non-vanishing frequency modes). Hence we have \emph{six} equations in total, but \emph{eight} variables. This issue of having more variables than equations is resolved by providing the equations of state
\begin{align}\label{sound}
\delta p_{r}=c_{s}^{(r)}\delta \rho~,
\qquad 
\delta p_{t}=c_{s}^{(t)}\delta \rho\,. 
\end{align}
which determines $\delta p_t$ and $\delta p_r$ as a function of $\delta \rho$. Choices for the sound speeds $c_{s}^{(r)}$ and $c_{s}^{(t)}$ close the system of equations.

We would like to point out that a similar set of equations have also been derived in \cite{Mondal:2023wwo}, albeit in the connection with the f-mode oscillation of anisotropic neutron stars. The parametrization of the metric and matter perturbations, considered in \cite{Mondal:2023wwo}, are different from those presented in this work, and hence the perturbation equations are not directly related with each other.

Given the curiosity that if there is a possibility to derive a second order master differential equation for polar perturbation with matter, alike the axial case, we first determine a system of two differential equations of first order that depends solely on the gravitational perturbations $K$ and $H_{1}$, describing the even perturbations, along with source terms. For that purpose, we have solved for $H(r)$ from Eq. \eqref{zerillimat06} and then have substituted the same in the differential equations for $K$, in Eq. \eqref{zerillimat01}, yielding, 
\begin{align}
K'=\left[\alpha_{0}(r)+\alpha_{2}(r)\omega^{2}\right]K+\left[\beta_{0}(r)+\beta_{2}(r)\omega^{2}\right]\left(\frac{H_{1}}{\omega}\right)+S_{\rm K}
\label{Kprimeling}
\end{align}
with the following definitions for the functions $\alpha_{0}(r)$, $\alpha_{2}(r)$, $\beta_{0}(r)$, $\beta_{2}(r)$, and $S_{\rm K}$ respectively, as (defining, $\gamma_{\ell}\equiv(\ell-1)(\ell+2)$ in subsequent expressions),
\begin{align}
\alpha_{0}&\equiv \frac{\gamma_{\ell}\left(1-g+8\pi r^{2}\bar{p}_{r}\right)}{2rg\left[\gamma_{\ell}+8\pi r^{2}\bar{p}_{r}+3(1-g)\right]}
-\frac{(1-g)\left(3g-1-8\pi r^{2}\bar{p}_{r}\right)}{rg\left[\gamma_{\ell}+8\pi r^{2}\bar{p}_{r}+3(1-g)\right]}~,
\quad
\alpha_{2}\equiv -\frac{2r}{f\left[\gamma_{\ell}+8\pi r^{2}\bar{p}_{r}+3(1-g)\right]}~,
\\
\beta_{0}&\equiv -\frac{\left(\gamma_{\ell}+2\right)\left\{\gamma_{\ell}+2(1-g)\right\}}{2ir^{2}\left[\gamma_{\ell}+8\pi r^{2}\bar{p}_{r}+3(1-g)\right]}~,
\qquad
\beta_{2}\equiv \frac{2ig}{f\left[\gamma_{\ell}+8\pi r^{2}\bar{p}_{r}+3(1-g)\right]}~,
\\
S_{\rm K}&=\frac{16\pi r\delta p_{r}}{\gamma_{\ell}+8\pi r^{2}\bar{p}_{r}+3(1-g)}-\frac{D^{\rm up}\left[\gamma_{\ell}+2(1-g)\right]}{2i\omega r^{2}\left[\gamma_{\ell}+8\pi r^{2}\bar{p}_{r}+3(1-g)\right]}~.
\end{align}
On the other hand, substitution of $H$ from Eq. \eqref{zerillimat06} in the differential equation for $H_{1}$, as in Eq. \eqref{zerillimat03}, yields the following first order differential equation for $H_{1}$,
\begin{align}
\frac{H_{1}'}{\omega}=\left[\kappa_{0}(r)+\kappa_{2}(r)\omega^{2}\right]K+\left[\delta_{0}(r)+\delta_{2}(r)\omega^{2}\right]\left(\frac{H_{1}}{\omega}\right)+S_{H1},
\label{Hprimeling} 
\end{align}
where, the coefficients $\kappa_{0}$, $\kappa_{2}$, $\delta_{0}$, $\delta_{2}$ and the source term $S_{H1}$ becomes,
\begin{align}
\kappa_{0}&\equiv -\frac{i\left[4\gamma_{\ell}g+16\pi r^{2}g\bar{p}_{r}+6g(1-g)-\left(3g-1-8\pi r^{2}\bar{p}_{r}\right)\left(g-1-8\pi r^{2}\bar{p}_{r}\right)\right]}{2g^{2}\left[\gamma_{\ell}+8\pi r^{2}\bar{p}_{r}+3(1-g)\right]}~,
\\
\kappa_{2}&\equiv \frac{2ir^{2}}{fg\left[\gamma_{\ell}+8\pi r^{2}\bar{p}_{r}+3(1-g)\right]}~,
\\
\delta_{0}&\equiv -\frac{1-g-4\pi r^{2}\left(\bar{\rho}-\bar{p}_{r}\right)}{rg}-\frac{\left(\gamma_{\ell}+2\right)\left(8\pi r^{2}\bar{p}_{r}+1-g\right)}{2rg\left[\gamma_{\ell}+8\pi r^{2}\bar{p}_{r}+3(1-g)\right]}~,\quad
\delta_{2}\equiv \frac{2r}{f\left[\gamma_{\ell}+8\pi r^{2}\bar{p}_{r}+3(1-g)\right]}~,
\\
S_{H1}&\equiv -\frac{4fC^{\rm up}}{g\omega}-\frac{16\pi i r^{2}\delta p_{r}}{g\left[\gamma_{\ell}+8\pi r^{2}\bar{p}_{r}+3(1-g)\right]}+\frac{D^{\rm up}(g-1-8\pi r^{2}\bar{p}_{r})}{2gr\omega\left[\gamma_{\ell}+8\pi r^{2}\bar{p}_{r}+3(1-g)\right]}~.
\end{align}
It is interesting to point out that we have the following useful relations:
\begin{align}
\delta_2=-\alpha_2,\qquad \kappa_2 = -\frac{\alpha_2^2}{\beta_2}~.     
\end{align}
Therefore, instead of the three equations, namely Eq.\eqref{zerillimat01}, Eq.\eqref{zerillimat03} and Eq.\eqref{zerillimat06} one can simply use the system of two first order equations presented in Eqs. \eqref{Kprimeling} and \eqref{Hprimeling}. This corresponds to the master system of equations for the even perturbations of a spherically symmetric metric. 

On the other hand, if we demand the existence of a single master function $\Psi_{\rm Z}$, such that,
\begin{align}
K&=\alpha(r)\Psi_{\rm Z}+\beta(r)\dfrac{d\Psi_{\rm Z}}{dr_{*}}~,
\label{zerillitest1g}
\\
\frac{H_{1}}{\omega}&=\kappa(r)\Psi_{\rm Z}+\delta(r)\dfrac{d\Psi_{\rm Z}}{dr_{*}}~,
\label{zerillitest2g}
\end{align}
where, the tortoise coordinate $r_{*}$ is defined as $(dr/dr_*)\equiv \sqrt{f g}$, and, if we further demand that $\Psi_{\rm Z}$ must satisfy an equation of the form
\begin{align}
\dfrac{d^{2}\Psi_{\rm Z}}{dr_{*}^{2}}=\left[V_{\rm Z}-\omega^{2}\right]\Psi_{\rm Z}+S_{Z}~,
\end{align}
we obtain an obstruction as long as $f \partial_r g - g \partial_r f \neq 0$. This obstruction disappears in the Schwarzschild case, where $f=g$ and hence one can define the Zerilli function and its associated potential. We refer the reader to  \ref{app:Schw}, where the Schwarzschild case is reviewed. Unfortunately, in the presence of matter, there is no generalization of the Zerilli master function for generic even perturbations. We have also investigated whether a third order master equation of the form 
\begin{align}\label{badZ}
\dfrac{d^{2}\Psi_{\rm Z}}{dr_{*}^{2}}+[A_Z + B_Z \omega^2]\dfrac{d\Psi_{\rm Z}}{dr_{*}}=\left[V_{\rm Z}-\omega^{2}\right]\Psi_{\rm Z}+S_{Z}~,
\end{align}
would exist and found a negative result, which contradicts the claims of \cite{Liu:2022csl}. For clarity, the third order term is $\omega^2(d\Psi_{\rm Z}/dr_{*})$, which reads in time domain as $(\partial^3\Psi_{\rm Z}/\partial t^2 \partial r_{*})$. Our proof amounts to show that Eqs. \eqref{zerillitest1g} - \eqref{zerillitest2g} and \eqref{badZ} lead to a contradiction as long as $f \partial_r g - g \partial_r f \neq 0$, or equivalently, as long as $f$ is not proportional to $g$. The proof is given in a Mathematica sheet in this \href{https://github.com/gcompere/Tidal-Love-numbers-and-quasi-normal-modes-in-the-Schwarzschild-Hernquist-dark-matter-halo}{Github} repository. Our result implies that the third order master equations described in \cite{Liu:2022csl} are gauge dependent.

\subsubsection{Zero frequency modes}\label{zero_polar}

Let us study the static modes, which are defined as the modes with $\omega=0$ in the frequency domain, and are important for the study of TLNs. We will concentrate on the $\ell \geq 2$ modes, as our interest is in the gravitational perturbations. First of all, the `$\theta$' component of the matter equations give
\begin{align}
\delta p_t  = \frac{\bar \rho+\bar p_r}{2} H\,. \label{deltapt}
\end{align}
The $(t,r)$ components of the perturbed Einstein equations imply that
\begin{align}
H_{1}=-\frac{1}{\ell(\ell+1)}D^{\rm up}~.
\end{align}
The $(t,\theta)$ components of the perturbed Einstein equations or, equivalently, the $t$ component of the perturbed matter equations then relates two matter perturbations as 
\begin{align}\label{CupDup}
\ell(\ell+1) C^{\rm up} = \frac{g}{4f}D^{\rm up \prime}+\frac{1-g+4\pi r^2(\bar p_r -\bar \rho)}{4 r f}D^{\rm up}. 
\end{align}
We solve these three algebraic equations for $\delta p_{\rm t}$, $H_1$ and $C^{\rm up}$. Note that the `$t$' component of the perturbed matter conservation equation, i.e., Eq. \eqref{conservpertt}, in the static limit, does not provide any new equation, rather is identical to Eq. \eqref{CupDup}. Finally, the `$r$' component of the perturbed matter conservation equation, presented in Eq. \eqref{conservpertr}, in the static limit, allows us to solve for $\delta \rho=\delta \rho[K',H,H',\delta p_r,\delta p_r']$. Explicitly, the corresponding expression reads,  
\begin{align}\label{conservpertrstatic}
\left(\frac{f'}{2f}\right)\delta \rho&=-\left(\frac{1+3g+8\pi r^{2}\bar{p}_{r}}{2rg}\right)\delta p_{r}+\left(\frac{\bar{p}_{r}+\bar{\rho}}{r}\right)H-\delta p_{r}'+\left(\frac{\bar{p}_{r}+\bar{\rho}}{2}\right)H'
\nonumber
\\
&+\left[\frac{r\bar{p}_{r}'}{2}+\frac{(\bar{\rho}+\bar{p}_{r})(1-g+8\pi r^{2}\bar{p}_{r})}{4g}\right]K'~.
\end{align}
Thus we have exhausted all the equations arising from the matter sector. 

Turning now, to the perturbations equations from the gravitational sector, we first express the static limit of the $(r,\theta)$ and the $(r,r)$ components, yielding two first order differential equations for $H$ and $K$,
\begin{align}
K'&-H'+\frac{g-1-8\pi r^2 \bar p_r}{r g}H=0\,,
\label{Kp}
\\
H'&-\frac{1+g+8\pi r^2 \bar p_r}{2g}K'-\frac{(\ell-1)(\ell+2)-16\pi r^2 \bar p_r}{2r g}H+\frac{(\ell+2)(\ell-1)}{2 r g}K=-\frac{8\pi r}{g}\delta p_r\,. 
\label{Hp}
\end{align}
One can first algebraically solve Eq. \eqref{Kp} for $K'$, substitute in Eq. \eqref{Hp} and solve for $K$. This yields,
\begin{align}\label{Kstatic}
K=\left[\frac{1-g+8\pi r^{2}\bar{p}_{r}}{(\ell-1)(\ell+2)}\right]\left(rH'\right)+\left[\frac{(\ell-1)(\ell+2)g-16\pi g r^{2}\bar{p}_{r}+(1+8\pi r^{2}\bar{p}_{r})^{2}-g^{2}}{(\ell-1)(\ell+2)g} \right]H-\frac{16 \pi  r^{2}}{(\ell-1)(\ell+2)}\delta p_{r}~.
\end{align}
The compatibility of the solutions for $K'$ and $K$, along with the `$r$' component of the perturbed matter conservation equation, then leads to a second order differential equation for the metric perturbation $H(r)$, yielding,
\begin{align}\label{Hdiffstatic}
H''&+\frac{1+g+4\pi r^2 (\bar p_r-\bar\rho)}{rg}H'+H\Bigg[\dfrac{d}{dr}\left(\frac{1-g+8\pi r^{2}\bar{p}_{r}}{rg}\right)+\frac{(1-g+8\pi r^2 \bar p_r)\{2g-4\pi r^2 (\bar{\rho}+\bar p_r)\}}{r^{2}g^{2}}
\nonumber
\\
&-\frac{\ell(\ell+1)-8\pi r^{2}(\bar{\rho}+\bar p_r)}{r^{2}g}\Bigg]+\frac{8\pi (\delta \rho+\delta p_{r})}{g}=0~.
\end{align}
The above differential equation for $H(r)$ can also be obtained via the following two routes --- (a) $g\times (t,t)$ component +$(\theta,\theta + \phi,\phi)$ component + $(g/r)\times (r,r)$ component, as well as (b) $(t,t)$ component + $(1/rg)\times$$(r,r)$ component. Note that both of these uses the second order differential equations for $H$ and for $K$. Of course, Eq. \eqref{Kp} can be used to eliminate any $K'$ in the above equations. It turns out that both of these methods yield the above equation, where Eq. \eqref{deltapt} needs to be used. 

Note that we still need a relation between $\delta \rho$ and $\delta p_{r}$, in order to close the above equations. However, then one needs to solve the second order equation for H, given by Eq. \eqref{Hdiffstatic} and the first order equation for $\delta \rho$, given by Eq. \eqref{conservpertrstatic}, together. In order to close the system of equations, given that we are interested in a DM halo distribution, we now assume that the radial sound speed $c_{s}^{(r)}$ is identically zero, which implies $\delta p_{\rm r}=0$. For simplicity of this presentation and consistency with the DM profile considered in this paper, we further take $\bar p_{\rm r}=0$ from now on. In which case, we obtain the following equation for $\delta \rho$, in terms of $H$ and $H'$,
\begin{align}
\frac{\delta \rho}{\bar{\rho}}=\left(\frac{1-2g+5g^{2}}{2(1-g)g}\right)H+\left(\frac{r(1+g)}{2(1-g)}\right)H'~.
\end{align} 
Note that the above equation can also be expressed in terms of $K$ and $H$ in the following manner:
\begin{align}
\delta \rho = \frac{(\ell+2)(\ell-1)g (\bar \rho+2 \bar p_t)}{(g-1)^2}K-\frac{\bar\rho+g^2(\bar\rho-4 \bar p_t)+g\{2\ell (\ell+1)\bar p_t +(\ell^2+\ell-4)\bar\rho\}}{(g-1)^2}H~.
\label{deltarho}
\end{align}
To see the equivalence between these two equations for $\delta \rho$, one simply needs to use the relation, $2\bar{p}_{t}=\{(1-g)\bar{\rho}/2g\}$. Substituting the above solution for $\delta \rho$, along with $\bar{p}_{r}=0=\delta p_{r}$, in Eq. \eqref{Hdiffstatic}, we obtain the final differential equation for the metric perturbation $H$,
\begin{align}\label{omega0evenv1}
H''+\left(\frac{1+g}{r g}+\frac{16\pi r \bar p_t}{1-g}+\frac{4\pi r(1-3g)\bar \rho}{g(g-1)}\right)H'+\left( -\frac{1+(\ell+2)(\ell-1)g+g^2}{r^2 g^2}+\frac{32\pi \bar p_t}{g}+\frac{16 \pi \bar\rho}{g(1-g)}\right)H=0~, 
\end{align}
which can also be expressed in the following suggestive form using background Einstein's equations:
\begin{align}\label{omega0even}
H''+\left(\frac{1+g}{r g}+\frac{8\pi r\bar{\rho}}{1-g}\right)H'+\left(-\frac{1+(\ell+2)(\ell-1)g+g^2}{r^2 g^2}+\frac{8\pi \bar{\rho}(1+g^{2})}{g^{2}(1-g)}\right)H=0~.
\end{align}
This is the master equation for static even modes in the case $\bar p_r=0=\delta p_r$, which we will solve to find out the electric type TLNs associated with the gravitational perturbation of a non-rotating BH immersed in a DM halo. We now move on to the discussion of the two observables, the QNMs and the TLNs. 

\section{Quasi-normal modes of the Schwarzschild-Hernquist BH}

Having presented all the relevant equations associated with both the matter and the gravitational perturbations, in this section we will discuss one of the observable associated with perturbation of a BH spacetime, namely the quasi-normal modes (QNMs). We will discuss the QNMs in the odd sector only. We will compare their values for the non-relativistic Hernquist profile versus the relativistic Schwarzschild-Hernquist profile. 

In the odd sector, one can introduce a Regge-Wheeler-like master function, and hence the master equation for the odd sector can be reduced to the following form 
\begin{align}\label{master_eq_odd}
\dfrac{d^{2}{\Psi}_{\rm RW}}{dr_{*}^{2}}+\left[\omega^{2}-V_{\rm RW}^{\rm up/down}\right]{\Psi}_{\rm RW}
=0~, 
\end{align}
where, $\Psi_{\rm RW}=(\sqrt{fg}/r)h_{1}$, and the expressions for the potentials $V_{\rm RW}^{\rm up/down}$ have been presented in Eqs. \eqref{RW_Pot_Up} and \eqref{RW_Pot_Down}, respectively. The asymptotic solutions to this equation take the form ${\Psi}_{\rm RW}\sim \exp(\pm i \omega r_*)$ and correspond to modes propagating outwards and inwards at spatial infinity and at the horizon. The quasi-normal mode eigenvalue problem is defined by requiring the modes to be purely outgoing at infinity and purely ingoing at the horizon. Explicitly, the boundary conditions read 
\begin{equation}
\Psi_{\rm RW}\sim\begin{cases}
e^{i\omega r}, \,\, r\rightarrow\infty, \\ (r-r_{\rm H})^{-i \omega r_{\rm H} }, \,\, r\rightarrow r_{\rm H}\,.
\end{cases}
\end{equation}
Here, $r_{\rm H}$ denotes the location of the event horizon. Once such boundary conditions are imposed, Eq. \eqref{master_eq_odd} admits a discrete set of complex eigenvalues, which are referred to as the QNMs. 

\begin{figure}[h]
\centering
\begin{subfigure}{.45\textwidth}
\captionsetup{singlelinecheck=false} 
\includegraphics[width=.95\linewidth]{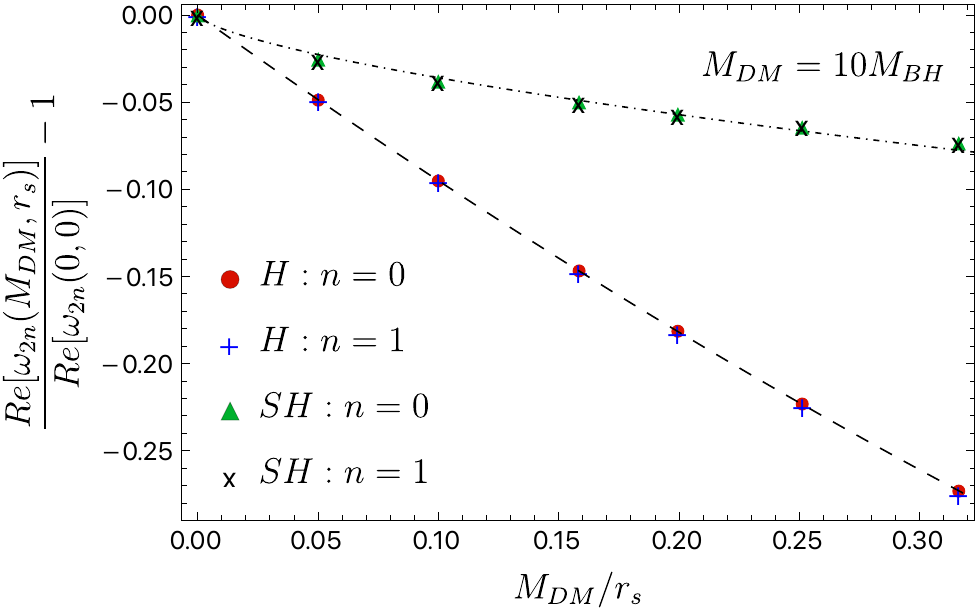}
\caption{Shift of the real part of the QNM frequency for $M_{\rm DM}=10\mbh$.}
\end{subfigure}
\quad 
\begin{subfigure}{.45\textwidth}
\captionsetup{singlelinecheck=false} 
\includegraphics[width=.95\linewidth]{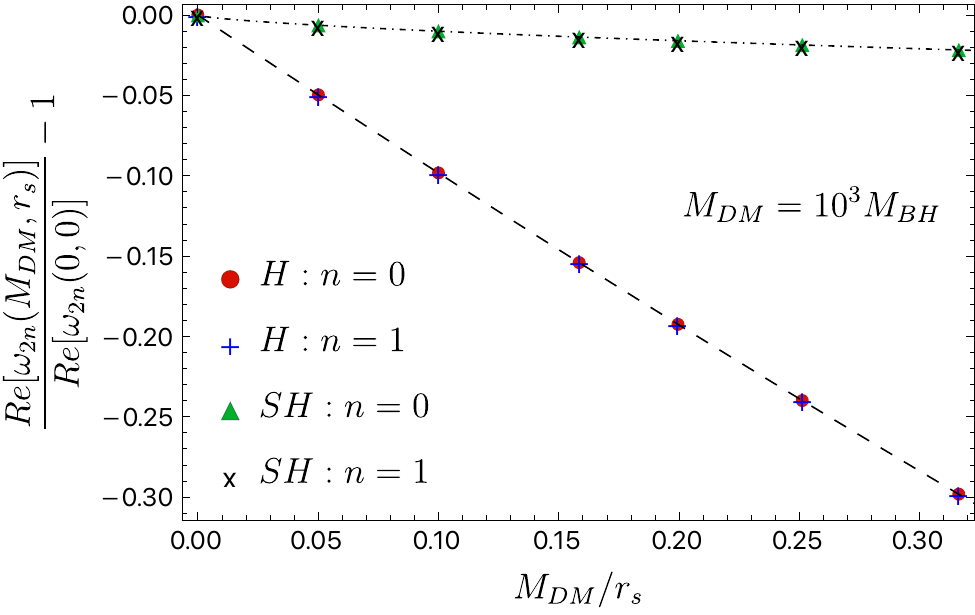}
\caption{Shift of the real part of the QNM frequency for $M_{\rm DM}=10^3\mbh$.}
\end{subfigure}
\vskip 0.5 cm
\begin{subfigure}{.45\textwidth}
\captionsetup{singlelinecheck=false} 
\includegraphics[width=.95\linewidth]{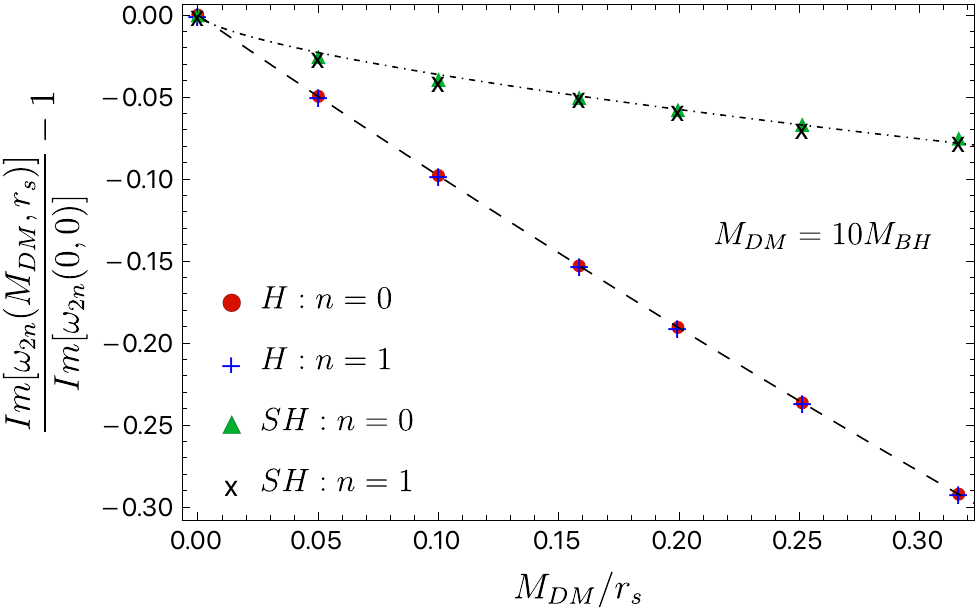}
\caption{Shift of the imaginary part of the QNM frequency for $M_{\rm DM}=10\mbh$.}
\end{subfigure}
\quad 
\begin{subfigure}{.45\textwidth}
\captionsetup{singlelinecheck=false} 
\includegraphics[width=.95\linewidth]{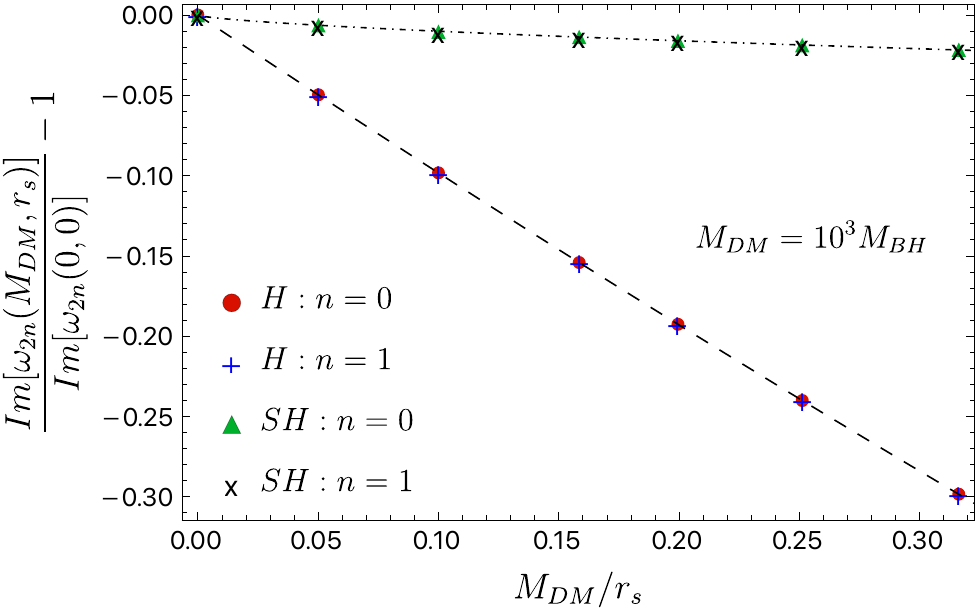}
\caption{Shift of the imaginary part of the QNM frequency for $M_{\rm DM}=10^3\mbh$.}
\end{subfigure}
\caption{Shifts in the $\ell=2$ mode frequency for the fundamental ($n=0$) and the first overtone ($n=1$) for both the Hernquist and Schwarzschild-Hernquist halos as a function of $M_{DM}/r_s$: (a) real part shift with $M_{DM}=10\mbh$; (b) real part shift with $M_{DM}=10^3\mbh$; (c) imaginary part shift with $M_{DM}=10\mbh$; (d) imaginary part shift with $M_{DM}=10^3\mbh$. The dashed lines are the best linear fits for the Hernquist model; the dot-dashed lines are the best fits for the Schwarzschild-Hernquist model.}
\label{QNMs_Shifts_overtones}
\end{figure}

Following the boundary conditions and the master equation for the axial sector, we have solved numerically for the QNM frequencies for the $\ell =2$ mode using a sixth-order WKB approximation method \cite{Konoplya:2003ii}. We compared the QNM frequencies derived from the Schwarzschild-Hernquist model, presented in \ref{sec:SH}, to the ones obtained from the analytical model derived in \cite{Cardoso:2021wlq} based on the Hernquist density profile.  
Both the Hernquist and the Schwarzschild-Hernquist models for DM only depend upon two dimensionless parameters $(M_{\rm DM}/M_{\rm BH})$, and $(r_{\rm s}/M_{\rm BH})$. In \ref{QNMs_Shifts_overtones} we have plotted the relative shift of the $\ell=2$ QNM frequencies $\omega_{2n}(M_{\rm DM},r_{\rm s})$, with respect to the Schwarzschild value $\omega_{2n}(0,0)$ for the fundamental mode $n=0$ and for the first overtone $n=1$. The plotted values were obtained from the 'down'  perturbation definition. For the Hernquist DM halo, the frequency shifts are linear in the dimensionless ratio $(M_{\rm DM}/r_{\rm s})$, i.e.
\begin{equation}
\frac{\omega_{2n}(M_{DM},r_s)}{\omega_{2n}(0,0)}-1=-\frac{M_{DM}}{r_s}+\order\left(\frac{M_{DM}^2}{r_s^2}\right)\,.
\end{equation}
This results can be understood in terms of the correction to the light ring frequency, which in the WKB approximation relates directly to the quasi-normal mode frequency, see e.g. \cite{Cardoso:2008bp,Cardoso:2021wlq} for more details. 

In the case of the Schwarzschild-Hernquist distribution, the shifts can instead be fitted with a power law 
\begin{equation}\label{schhern_qnm}
\frac{\omega_{2n}(M_{DM},r_s)}{\omega_{2n}(0,0)}-1=-c\left(\frac{M_{DM}}{r_s}\right)^{3/2}+\order\left(\frac{M_{DM}^2}{r_s^2}\right)\,,
\end{equation}
with, $c=c(M_{\rm DM})$ is some mass-dependent coefficient. We have numerically derived that $c$ can be well approximated as $c(M_{\rm DM})=1/(2\,M_{\rm DM}^{1/3})$ over a large parameter range. This, along with the scaling in Eq. \eqref{schhern_qnm}, are one of the main results of this paper. 

\begin{figure}[h]
\centering
\begin{subfigure}{.45\textwidth}
\captionsetup{singlelinecheck=false} 
\includegraphics[width=.95\linewidth]{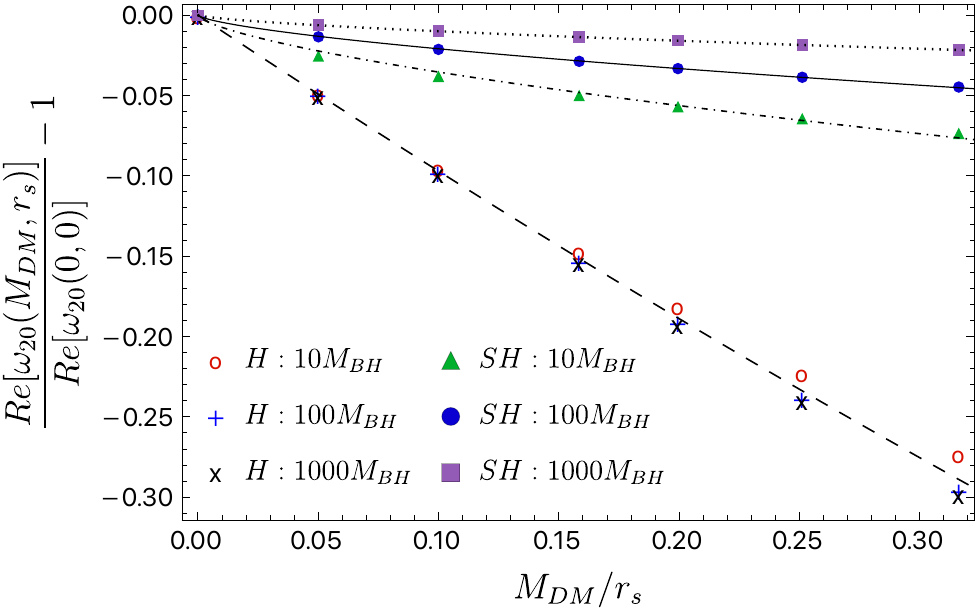}
\caption{Relative shift in the real part of the QNM frequency compared to the Schwarzschild BH.}
\end{subfigure}
\begin{subfigure}{.45\textwidth}
\captionsetup{singlelinecheck=false} 
\includegraphics[width=.95\linewidth]{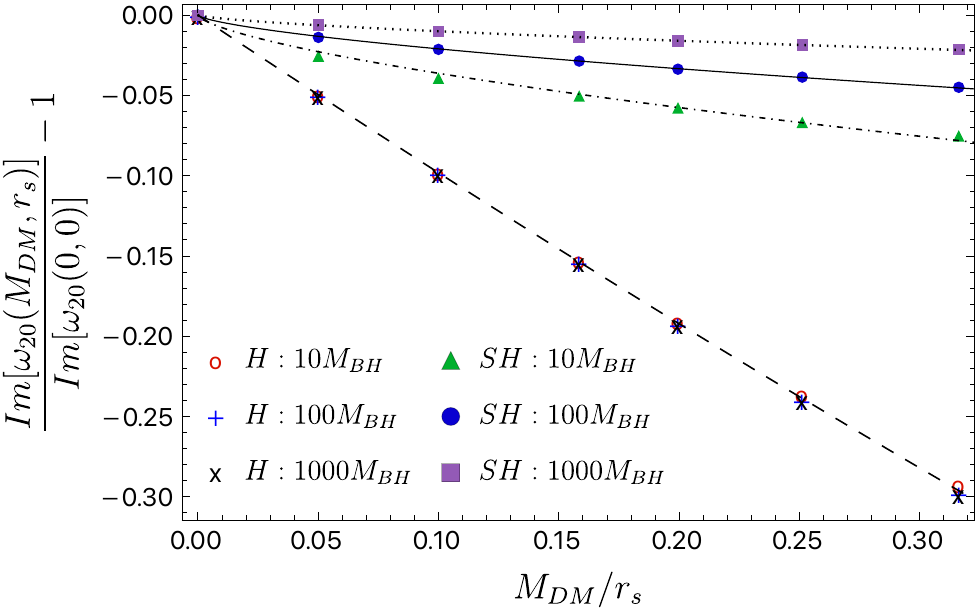}
\caption{Relative shift in the imaginary part of the QNM frequency compared to the Schwarzschild BH.}
\end{subfigure}
\caption{The shift in the $\ell=2$, $n=0$ mode frequency for the Hernquist and Schwarzschild-Hernquist DM halos as a function of the dimensionless ratio $(M_{\rm DM}/r_{\rm s})$: (a) relative real part shift; (b) relative imaginary part shift, both with respect to their Schwarzschild value.}
\label{QNMs_Shifts_fundamental}
\end{figure}

To summarize, in the case of Hernquist DM density profile, the shifts in the QNM frequency show a universal scaling with the ratio $(M_{\rm DM}/r_{\rm s})$, whereas in the case of Schwarzschild-Hernquist profile, the scaling with $(M_{\rm DM}/r_{\rm s})$ is different, and in addition it is weighted by the mass-dependent term $c(M_{\rm DM})$. This is summarised in \ref{QNMs_Shifts_fundamental}. It is important to mention that the same scaling emerges while studying the variation to the light ring frequency $\Omega_{\rm LR}\equiv (\sqrt{f(r_{\rm LR})}/r_{\rm LR})$ as a function of the Schwarzschild-Hernquist halo parameters. Intriguingly, the above scaling holds true for both the real as well as the imaginary part of the QNM frequency. 

\begin{figure}[h]
\centering
\begin{subfigure}{.45\textwidth}
    \captionsetup{singlelinecheck=false} 
\includegraphics[width=.95\linewidth]{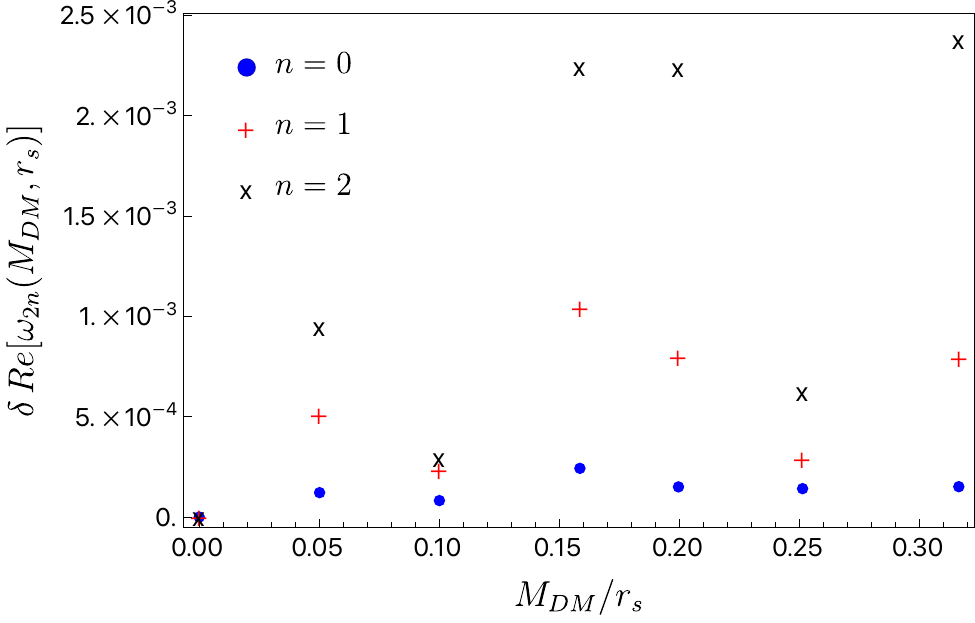}
\caption{Relative difference in the real part of the QNM frequency for $M_{\rm DM}=10\mbh$.}
\end{subfigure}
\quad
\begin{subfigure}{.45\textwidth}
\captionsetup{singlelinecheck=false} 
\includegraphics[width=.95\linewidth]{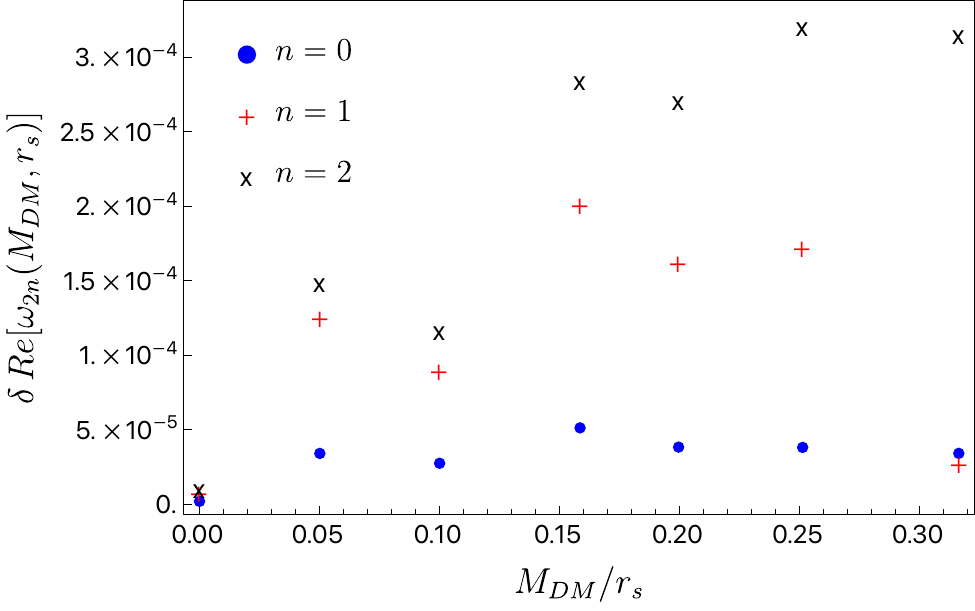}
\caption{Relative difference in the real part of the QNM frequency for $M_{\rm DM}=100\mbh$.}
\end{subfigure}
\caption{Difference in the real part of the $\ell=2$ mode frequencies between the `up' and the `down' sector for the fundamental ($n=0$), first ($n=1$) and second ($n=2$) overtone for the Schwarzschild-Hernquist DM halo profile as a function of the dimensionless ratio $(M_{\rm DM}/r_{\rm s})$ for different DM masses:  (a) $M_{\rm DM}=10\mbh$, (b) $M_{\rm DM}=100\mbh$.}
\label{QNMs_Shifts_UpDown}
\end{figure}

We showed in \ref{perturbation_odd} that the effective potential of the Regge-Wheeler equation depends upon how the fluid four-velocity is being perturbed. Depending on whether, $\delta u^{\mu}=0$, or, $\delta u_{\mu}=0$, we have two different effective potentials, given by Eqs. \eqref{RW_Pot_Up} and \eqref{RW_Pot_Down}. This suggests that these two definitions will give distinct QNM spectrum. However, we have verified numerically that the relative difference between the QNM frequencies in the 'up' and the 'down' definitions is smaller than $\order(10^{-4})$, over the parameter space considered in this work. Thus, the QNM frequencies computed with the two potentials coincide up to numerical uncertainty. This feature can be see in \ref{QNMs_Shifts_UpDown}, where we have depicted the fundamental mode and the first overtone computed with the `up' and the `down' potentials for different masses of the DM halo.

Let us finally mention how one could obtain the QNMs in the even sector. In \ref{sec:even} we have obtained four coupled ordinary differential equations for the metric perturbations $K$, $H_1$, and the matter perturbations $\delta\rho$ and $D^{\rm up}$. In the range $2\mbh<r<4\mbh$, since there is no matter, one can use Eqs. \eqref{zerillitest1g} and \eqref{zerillitest2g} to solve for the Zerilli master equation. The solution, which is ingoing at the horizon, is well-known in terms of Heun functions,
using which we deduce the following boundary condition for the functions $K$, $H_1$, $\delta\rho$ and $D^{\rm{up}}$ at $4\mbh$: $K(4\mbh)$ and $H_1(4\mbh)$ are deduced from the Zerilli ingoing function, while $\delta\rho(4\mbh)=0$ and $D^{\rm up}(4\mbh)=0$. We then use these boundary conditions to evolve the system of four ordinary first order differential equations up to a radial cutoff $r_{\rm C}$. At large radius, we can assume that the matter content is negligible and use the known 'up' solution of the Zerilli equation with outgoing boundary conditions at infinity. We deduce from Eqs. \eqref{zerillitest1g} and \eqref{zerillitest2g} the resulting boundary conditions for $K(r_{\rm C})$ and $H_1(r_{\rm C})$. In order to have a well-posed system we keep the boundary conditions $D^{\rm{up}}(r_{\rm C})$ and $\delta\rho(r_{\rm C})$ as free parameters. We then evolve the system inwards using the four ordinary first order differential equations. We perform a shooting method, where we equate the solutions at an intermediate radius between $4\mbh$ and $r_{\rm C}$, fixing the real and imaginary parts of the frequency as well as the boundary conditions $D^{\rm{up}}(r_{\rm C})$ and $\delta\rho(r_{\rm C})$. There is a discrete set of solutions which are the frequencies of the QNMs.

\section{Tidal Love numbers of a black hole in a dark matter halo}

In this section, we consider the BH-spike system as a self-gravitating object subject to an external tidal field. We study the tidal Love numbers (TLNs) of the original non-relativistic Hernquist profile and compare them with the relativistic Schwarzschild-Hernquist profile described earlier in this work. The external tidal field can be expanded in polar and axial multiple moments, with amplitudes $\mathcal{E}^{\ell m}$ and $\mathcal{B}^{\ell m}$ when projected on a spherical harmonics basis $Y^{\ell m}(\theta,\phi)$ \cite{Binnington:2009bb}. As a result of this interaction, the mass and current multipole moments of the system $(M_\ell,S_\ell)$ change proportionally to the external field. The system multipole moments can be obtained from the asymptotic expansion of the metric functions, via

\begin{align}
    g_{tt}&=-1+\frac{2M}{r}+\sum_{\ell\geq 2}\left( \, \frac{2}{r^{\ell+1}}\left[\sqrt{\frac{4\pi}{2\ell+1}}M_\ell\,Y^{\ell0}+\left(\ell'<\ell \, \text{pole}\right)\right]-\frac{2}{\ell\left(\ell-1\right)}r^\ell\left[\mathcal{E}_\ell Y^{\ell0} +\left(\ell'<\ell \, \text{pole}\right) \right]   \right) \,,\\
	g_{t\varphi}&=\frac{2J}{r}\sin^2\theta+\sum_{\ell\geq 2}\left( \, \frac{2}{r^{\ell}}\left[\sqrt{\frac{4\pi}{2\ell+1}}\frac{S_\ell}{\ell}\,S^{\ell0}_\varphi+\left(\ell'<\ell \, \text{pole}\right)\right]+\frac{2r^{\ell+1}}{3\ell\left(\ell-1\right)}\left[\mathcal{B}_\ell S^{\ell0}_\varphi +\left(\ell'<\ell \, \text{pole}\right) \right]   \right) \, ,\label{eq:AxialMetricExpansion}
\end{align}
with $S^{\ell m}=\sin(\theta)Y^{\ell m}_{,\theta}$. We can then define the electric (even) and magnetic (odd) tidal Love numbers as \cite{Hinderer:2007mb,Cardoso:2017cfl,Cardoso:2019upw},

\begin{align}
    k_\ell^E&=-\frac{1}{2}\frac{\ell\left(\ell-1\right)}{R^{2\ell+1}}\sqrt{\frac{4\pi}{2\ell+1}}\frac{M_\ell}{\mathcal{E}_{\ell}} \, , \\
	k_\ell^B&=-\frac{3}{2}\frac{\ell\left(\ell-1\right)}{\left(\ell+1\right)R^{2\ell+1}}\sqrt{\frac{4\pi}{2\ell+1}}\frac{S_\ell}{\mathcal{B}_{\ell}}  \,  ,
\end{align}
with $R$ a characteristic scale of the system, for example the stellar radius in the neutron star case. We present our results for both the axial as well as the polar sector. In the axial sector, there are two different equations obtained from the `up' and the `down' choices of perturbations mentioned earlier, and we discuss how the TLNs differ for both of these cases. Moreover, for the non-relativistic profile, the TLNs can be determined analytically up to linear order in the ratio of DM mass and halo radius, while for the relativistic profile we will use a robust numerical method for determining the TLNs. We start with the axial sector. 

\subsection{Odd sector}

Our starting point is the static limit $\omega=0$ of the linearized perturbations of Einstein's equations. As discussed in \ref{perturbation_odd}, we found that there are two natural definitions for the matter sector perturbations: either one can fix $\delta (u^\mu)=0$, which we refer to as the 'up' definition, or, one can fix $\delta (u_\mu)=0$ which we refer to as the 'down' definition. Correspondingly, the master equation for the axial variable $h_{0}$ (associated with the $\delta g_{t\phi}$ component) in the static limit is given by setting $\omega=0$ in either Eq. \eqref{eq3} (corresponding to the 'up' case) or Eq. \eqref{eq3bis} (corresponding to the 'down' case). We will discuss these two cases separately for both relativistic and non-relativistic DM profiles. 
\subsubsection{The 'up' axial TLNs for the non-relativistic Hernquist profile}
\label{upaxialNR}

Using the 'up' definition of the fluid perturbations, and in the static limit, we find the metric perturbation $h_{0}$ to satisfy the following second order differential equation: 
\begin{equation}\label{eq3bisstaticUP}
\qquad g h_0'' -4\pi r\left(\bar p_{\rm r}+\bar \rho\right)h_0'-\left[\frac{(\ell+2)(\ell-1)}{r^2}+\frac{2g}{r^2}+8\pi\left(\bar \rho-\bar p_{\rm r}+2 \bar p_{\rm t}\right)\right]h_0=0\,.
\end{equation}
Inserting the background fluid quantities, namely $(\bar \rho,\bar p_{\rm r},\bar p_{\rm t})$, it becomes challenging to obtain an analytic solution for the metric element $h_0$ in full generality. We thus expand $h_0$ in terms of the small parameter $\epsilon\sim (M_{\rm DM}/r_{s})$. For a realistic astrophysical scenario, with values of various scales given by Eq. \eqref{Ref1}, the above introduced smallness parameter $\epsilon$ becomes,
\begin{align}
\epsilon \equiv \frac{G M_{\rm DM}}{c^2 r_s} \sim 10^{-4}~.
\end{align}
As evident, the above quantity is indeed a small number and hence we can expand the relevant expressions in powers of the smallness parameter $\epsilon$. 
Following which, we may write $h_0=h_0^{(0)}+\epsilon h_0^{(1)}+\order(\epsilon^2)$, where the zeroth order term satisfies the standard equation for a static axial perturbation around the Schwarzschild BH,  
\begin{align}\label{eqh0Scd}
h^{(0)\prime\prime}_{0}&-\left[\frac{\ell(\ell+1)r-4\mbh}{r^{2}(r-2\mbh)}\right]h^{(0)}_{0}=0~.
\end{align}
Note that we have at hand an (even smaller) second small parameter 
\begin{align}
\eta \equiv \frac{G M_{\rm BH}}{c^2 r_s} \sim 10^{-12}~.
\end{align}
Focusing on the $\ell=2$ case, the solution to the above zeroth order equation takes the following form, 
\begin{align}\label{eqh0zerothorder}
h_{0}^{(0)}&=C_{1}r^{2}(r-2\mbh)+C_{2}\Big[\frac{2\mbh^{3}+2\mbh^{2}r+3\mbh r^{2}-3r^{3}}{12\mbh^{4}r}-\frac{r^{2}(r-2\mbh)}{8\mbh^{5}}\ln \left(1-\frac{2\mbh}{r}\right)\Big]~.
\end{align}
In order to have regularity of $h_0^{(0)}$ and its derivatives at the horizon, we must set $C_{2}=0$. Hence, we obtain, the following solution for the zeroth order metric perturbation in the axial sector: $h_{0}^{(0)}=C_{1}r^{2}(r-2\mbh)$. As a consequence the $\ell=2$ magnetic TLN of Schwarzschild BH vanishes, because there is no decaying mode in the solution for the response to the tidal perturbation. This is consistent with earlier findings in the literature.

Let us now turn our attention to the first order perturbation of the $\ell=2$ axial TLN due to the presence of DM halo. In the case of the non-relativistic Hernquist profile, we obtain the following equation for the metric perturbation, which is first order in the smallness parameter $\epsilon$,
\begin{align}
h^{(1)\prime\prime}_{0}&-\left[\frac{\ell(\ell+1)r-4\mbh}{r^{2}(r-2\mbh)}\right]h^{(1)}_{0}=\left[\frac{2r_s(r_s+2\mbh)}{(r+r_s)^{3}}\right]h^{(0)\prime}_{0}
\nonumber
\\
&\qquad +\left[\frac{2r_s(\ell-1)(\ell+2)}{r(r+r_s)^{2}}+\frac{4r_s(r_s+2\mbh)(r-\mbh)}{r(r+r_s)^{3}(r-2\mbh)}\right]h^{(0)}_{0}~.
\end{align}
Therefore, the zeroth order term acts as a source for the differential equation satisfied by the first order term. In the case of $\ell =2$, substituting $h_{0}^{(0)}=C_{1}r^{2}(r-2\mbh)$, the above differential equation for the first order perturbation $h_0^{(1)}$ becomes
\begin{align}
h^{(1)\prime\prime}_{0}&-\left[\frac{6r-4\mbh}{r^{2}(r-2\mbh)}\right]h^{(1)}_{0}=C_{1}\frac{2r_s r}{(r+r_s)^{3}}\left[4r^{2}+r\left(9r_s+2\mbh\right)-2\mbh(7r_s+6\mbh)\right]~.
\end{align}
The above differential equation can be solved and the function $h_0^{(1)}$ admits a lengthy analytical solution with two arbitrary constants. The solution involves a homogeneous part, which resembles Eq.\eqref{eqh0zerothorder}, with two arbitrary constants, $D_{1}$ describing the growing mode and $D_{2}$ describing the decaying mode, along with a particular solution dependent on $C_{1}$. Imposing regularity at the horizon and discarding the redundant growing mode of the homogeneous solution (setting $D_{1}=0$), we obtain a unique solution for $D_{2}$, proportional to $C_1$. The $\ell=2$ magnetic TLN is extracted from the ratio of the coefficients of the $\order(r^{-2})$ and $\order(r^3)$ terms in the asymptotic expansion of $h_0$, times a suitable numerical coefficient. For $\ell=2$, this overall numerical factor turns out to be $-(1/3)$. At leading order in $\epsilon$, we take $C_1$ as the leading order $r^3$ term of $h_0^{(0)}$ while we extract the $r^{-2}$ term from $h_0^{(1)}$. Instead of providing the full answer, we will now keep only the leading result in the expansion in the small number $\epsilon$, yielding
\begin{equation}
k^{\rm B(up)}_{\ell=2}= \frac{M_{\rm DM} r^4_s \left[5+12 \log (r_s/R)\right]}{3 R^5}~,
\end{equation}
where, we have introduced an effective radial scale $R$, in order to make the argument of the logarithmic term dimensionless. 

The characteristic length scale $R$, introduced above, can be defined in several ways. If we choose $R=(M_{\rm DM}+M_{\rm BH})$, then the above result will coincide with the axial TLN derived in \cite{Cardoso:2021wlq}. However, this choice is inconsistent with the perturbation scheme. Indeed, one would then have $k^{\rm B(up)}_{\ell=2}=O(\epsilon^{-4})$ instead of $k^{\rm B(up)}_{\ell=2}=O(\epsilon^{1})$. If we set $R=M_{BH}$ then the TLN would be $O(\epsilon \eta^{-5})$ which is inconsistent with the perturbative scheme. Consistently with the perturbative scheme, we normalize the TLN with respect to $r_s$. The TLN becomes $k^{\rm B(up)}_{\ell=2}= \frac{5}{3}\epsilon$, which scales correctly in $\epsilon$, and differs from \cite{Cardoso:2019upw,Cardoso:2021wlq}.

\subsubsection{The `down' axial TLNs: Non-relativistic Hernquist profile}

In the `down' definition for the fluid perturbations, the perturbation equation for $h_{0}$, in the static limit, reduces to 
\begin{equation}\label{eq3bisstaticDOWN}
\qquad g h_0'' -4\pi r\left(\bar p_{\rm r}+\bar \rho\right)h_0'-\left[\frac{(\ell+2)(\ell-1)}{r^2}+\frac{2g}{r^2}-8\pi\left(\bar \rho+\bar p_{\rm r}\right)\right]h_0=0\,.
\end{equation}
Following the discussion in the `up' case, here also we insert the background fluid quantities, e.g., $g(r)$, $\bar{\rho}$, $\bar{p}$ in the above equation and then expand $h_0$ to first order in $\epsilon$. This yields two differential equations, one for the zeroth piece $h_{0}^{(0)}$ and the other one for the first-order piece $h_{0}^{(1)}$. The differential equation satisfied by $h_0^{(0)}$ is identical to Eq. \eqref{eqh0zerothorder}, while, the first-order perturbation $h_0^{(1)}$ satisfies a modified equation:
\begin{align}
h^{(1)\prime\prime}_{0}-\left[\frac{\ell(\ell+1)r-4\mbh}{r^{2}(r-2\mbh)}\right]h^{(1)}_{0}=&\left[\frac{2r_s(r_s+2\mbh)}{(r+r_s)^{3}}\right]h^{(0)\prime}_{0}
\nonumber
\\
+&\left[\frac{2r_s\left\{(\ell-1)(\ell+2)(r+r_s)-2(r_s+2\mbh)\right\}}{r(r+r_s)^{3}}\right]h^{(0)}_{0}~.
\end{align}
Using the fact that the solution for $h_{0}^{(0)}$, which is regular at $r=2M_{\rm BH}$, reads $h_{0}^{(0)}=C_{1}r^{2}(r-2M_{\rm BH})$, along with setting $\ell=2$, the above differential equation for $h_0^{(1)}$ reduces to,
\begin{equation}
h^{(1)\prime\prime}_{0}-\left[\frac{6r-4\mbh}{r^{2}(r-2\mbh)}\right]h^{(1)}_{0}=C_{1}\frac{2r_s r}{(r+r_s)^{3}}\left[4r^{2}+r\left(5r_s-6\mbh\right)-8r_s\mbh\right]\,.
\end{equation}
Repeating the same analysis as in the `up' case, elaborated in \ref{upaxialNR}, we obtain the axial TLN for the $\ell=2$ mode, associated with the down sector to read
\begin{equation}
k^{\rm B(down)}_{\ell=2}=\frac{M_{\rm DM}r_s^{4}\left[1-4 \log(r_s/R)\right]}{3R^{5}}~.
\end{equation}
Here also $R$ is characteristic length scale of the system, which we choose to be $r_s$, consistently with the perturbative scheme which requires that the TLN scales as $\epsilon$. The fractional change in the $\ell=2$ axial TLN between the `up' and the `down' definition of perturbations is given by,
\begin{equation}
\Delta k^{\rm B}_{\ell=2}\equiv \frac{k^{\rm B (up)}_{\ell=2}-k^{\rm B(down)}_{\ell=2}}{k^{\rm B (up)}_{\ell=2}}
=\frac{4}{5}~,
\end{equation}
which is a $\mathcal{O}(1)$ quantity. Therefore, the difference between the axial TLNs for the `up' and the `down' definition of perturbations is significant.

\subsubsection{The 'up' axial TLNs for the Schwarzschild-Hernquist profile}

The static axial perturbations, described by $h_{0}$, are again given by Eq. \eqref{eq3bisstaticUP}, irrespective of the DM profile under consideration. In the case of a relativistic Schwarzschild-Hernquist profile, it is challenging to obtain analytical solutions for the metric element $h_0$, even at linear order in the smallness parameter $\epsilon$, due to complicated nature of the density profile and mass function. 
Since our interest for the determination of the TLNs is in the intermediate regime, we will consider the tidal effects up to a scale $r \ll r_s' \sim 10^{16} M_{\rm BH}$ for the range of parameters considered in Eq. \eqref{Ref1}. The DM density profile then reads:
\begin{equation}
\bar{\rho}=\lambda \,A\left(1-\frac{4\mbh}{r}\right)^w\left(\frac{\mbh}{r}\right)^q,
\end{equation}
where, estimations for $\lambda$, $w$ and $q$ have been given in \ref{sec:SH}. Integration of the above density profile will yield the mass function $m(r)$, which will be given by a hypergeometric function. 

Since the relativistic dimensionless parameter 
\begin{align}
 \epsilon_{R} \equiv \lambda A M_{BH}^2\sim 5 \times 10^{-8}\lambda    
\end{align} 
sets the global scale of the DM density profile, we shall expand the expressions to order $\order( \epsilon_{R} )$, neglecting terms of $\order( \epsilon_{R}^2 )$. Thus, we can expand the axial metric perturbation as, $h_0=h_0^{(0)}+ \epsilon_{R}  h_1^{(0)}+\order( \epsilon_{R}^2)$ and expand Eq. \eqref{eq3bisstaticUP} to the relevant order. For the $\ell=2$ mode, the zeroth order perturbation $h_0^{(0)}$ satisfies again Eq. \eqref{eqh0Scd}, which has the following regular solution: $h_{0}^{(0)}=C_{1}r^{2}(r-2M_{\rm BH})$. Substituting back this solution for the zeroth order term, along with the density profile and the mass function, the differential equation for $h_{0}^{(1)}$ can be obtained. Due to its long and complicated expression, we have not quoted it here.  

Given this differential equation, we have solved it numerically with the following boundary conditions: $h_{0}^{(1)}(4\mbh)=0$ and $\partial_{r}h_{0}^{(1)}(4\mbh)=0$. Both of which stem from the results that the DM profile ends at $r=4M_{\rm BH}$, and hence at $r=4\mbh$, the perturbations must coincide with the Schwarzschild counterpart. Similarly, one can also solve the differential equation in the intermediate zone through an asymptotic series, which involves $C_{1}$, as well as two additional constants $D_{1}$ and $D_{2}$. Subsequent matching of this asymptotic solution with the near-zone fixes the two constants $D_{1}$ and $D_{2}$ in terms of $C_{1}$, and hence one obtains the TLN associated with the $\ell=2$ mode as,
\begin{equation}
k^{\rm B(up)}_{\ell=2}\simeq -279.86\,\frac{M_\text{BH}^5}{R^5} \epsilon_{R}  ~.
\end{equation}
Consistently with the perturbation scheme, we set $R=M_\text{BH}$ as standard for the computation of the TLN of a black hole, so that the TLN is linear in $\epsilon_R$. We have also verified that the above numerical result for the TLN is stable with respect to change in the matching radius, which validates our numerical scheme. 

The above TLN associated with the relativistic profile is also of the same order as that of the non-relativistic profile, albeit with an overall negative sign.

\subsubsection{The 'down' axial TLNs for the Schwarzschild-Hernquist profile}

The same expansion can be used to solve Eq. \eqref{eq3bisstaticDOWN}, stemming from the 'down' perturbation definition. Here also the zeroth order term satisfies the axial perturbation equation for the Schwarzschild spacetime, while the first order term satisfies a complicated equation. One can solve this equation in the near zone with identical boundary conditions, $h_{0}^{(1)}(4\mbh)=0$ and $\partial_{r}h_{0}^{(1)}(4\mbh)=0$, and also solve the equation in the far zone. Matching of the solution and its derivative between the near and the far zone uniquely fixes the arbitrary constants and, after fixing $R=M_\text{BH}$ as above, yields the following TLN for the axial $\ell=2$ mode, 
\begin{equation}
k^{\rm B(down)}_{\ell=2}\simeq -102.1  \epsilon_R\,.
\end{equation}
Alike the case of `up' perturbation, for the 'down' perturbation as well the TLN is negative, and similar to the non-relativistic profile, for the relativistic case as well the relative difference in the TLN between the 'up' and 'down' perturbation schemes is of $\order(1)$, in particular,
\begin{equation}
\Delta k^{\rm B}_{\ell=2}\equiv \frac{k^{\rm B (up)}_{\ell=2}-k^{\rm B(down)}_{\ell=2}}{k^{\rm B (up)}_{\ell=2}} \approx 0.63\,.
\end{equation}
Thus the relative difference of the TLNs between these two definitions is of order unity, though individually the corresponding TLNs are small. This finishes our discussion involving TLNs of BHs immersed in both non-relativistic as well as relativistic DM profiles in the axial sector. We now turn our attention to the TLNs in the polar sector. 

\subsection{Even sector}

Having discussed the axial TLNs of BHs with a DM halo, associated with the $\ell=2$ mode, in the previous section, here we present the corresponding results for the even sector. The starting point, once again, is the $\omega\to0$ limit of the perturbation equations for the polar sector. As discussed in \ref{zero_polar}, in this limit, one can obtain a single second-order ODE for the metric perturbation $H(r)$, which we have presented in Eq. \eqref{omega0even}. As already emphasized, in the polar sector, there are no distinctions between the `up' and the `down' components. Thus we start by solving the corresponding equation for non-relativistic Hernquist DM profile and hence determining the TLNs. 

\subsubsection{Polar TLNs: Non-relativistic Hernquist profile}

Alike the axial sector, Eq. \eqref{omega0even} does not admit an explicit solution for the pressure and density associated with the non-relativistic Hernquist profile. Thus we proceed by expanding the metric perturbation $H$ as $H=H^{(0)}+\epsilon H^{(1)}+\order(\epsilon^2)$, where $\epsilon=(M_{\rm DM}/r_s)$, and solve it order by order. First of all we note that for the non-relativistic Hernquist profile, Eq. \eqref{omega0even} can be simplified to the following form,
\begin{align}
H''&+\left[\frac{2\left\{r-m(r)\right\}}{r\left\{r-2m(r)\right\}}+\frac{4\pi r^{2}\bar{\rho}}{m(r)}\right]H'
\nonumber
\\
&+\left[-\frac{\ell(\ell+1)r\left\{r-2m(r)\right\}+4m(r)^{2}}{r^{2}\left\{r-2m(r)\right\}^{2}}+\frac{8\pi r\bar{\rho}\left\{r^{2}-2rm(r)+2m(r)^{2}\right\}}{m(r)\left\{r-2m(r)\right\}^{2}}\right]H=0.
\end{align}
After substituting the background expressions for the mass function $m(r)$ and density $\bar{\rho}(r)$ and expanding in powers of $\epsilon$, the zeroth and first order terms satisfy the following differential equations
\begin{align}
H^{(0)\prime\prime}&+\frac{2(r-\mbh)}{r(r-2 \mbh)}H^{(0)\prime}-\frac{4\mbh^{2}-2\ell(\ell+1)r\,\mbh+\ell(\ell+1)r^2}{r^2 (r-2 \mbh)^2}H^{(0)}=0~,
\\
H^{(1)\prime\prime}&+\frac{2(r-\mbh)}{r(r-2 \mbh)}H^{(1)\prime}-\frac{4\mbh^{2}-2\ell(\ell+1)r\,\mbh+\ell(\ell+1)r^2}{r^2(r-2 \mbh)^2}H^{(1)}
\nonumber
\\
&=\frac{2r_s\left[\mbh (r_s+4 \mbh)-r (r_s+3 \mbh)\right]}{\mbh (r_s+r)^3}H^{(0)\prime}
\nonumber
\\
&\qquad +\frac{2 r_s \left[r_s \ell(\ell+1) \mbh-2 r_s r+\mbh \left\{\left(\ell^2+\ell-4\right) r+4 \mbh\right\}\right]}{\mbh r(r_s+r)^3}H^{(0)}~.
\end{align}
Alike the axial TLNs, here also we focus on the $\ell=2$ mode. At the leading order in $\epsilon$, the differential equation for $H^{(0)}$ can be solved analytically and the solution reads,
\begin{equation}
H^{(0)}= C_1 \frac{3 r (r-2 \mbh)}{\mbh^2}+C_2\frac{ \mbh(r-\mbh) \left(2 \mbh^2+6 \mbh r-3 r^2\right)+\frac{3}{2} r^2 (r-2 \mbh)^2 \ln \left(1-\frac{2\mbh}{r}\right)}{\mbh^2 r (2 \mbh-r)}~,
\end{equation}
where $C_{1}$ and $C_{2}$ are arbitrary constants of integration. The regularity of the solution at the horizon requires that we set $C_{2}$ to zero and hence the solution reduces to $H^{(0)}= C_1 3 r (r-2 \mbh)/\mbh^2$. Substituting the above regular solution for $H^{(0)}$, the differential equation for $H^{(1)}$ reduces to, 
\begin{align}
H^{(1)\prime\prime}&+\frac{2(r-\mbh)}{r(r-2\mbh)}H^{(1)\prime}+\frac{2 \left(2 \mbh^2-6 \mbh r+3 r^2\right)}{r^2 (r-2 \mbh)^2}H^{(1)}
\nonumber 
\\
&=C_{1}\frac{12 r_s \left[\mbh^2 (7 r_s+8 \mbh)+2 r^2 (r_s+\mbh)-7 \mbh r (r_s+\mbh)\right]}{\mbh^3
(r_s+r)^3}\,.
\end{align}
It follows from the above equation that $H^{(1)}$ admits an analytic solution, which we have not presented here due to its complicated nature. The solution has two constants $D_{1}$, associated with growing mode, and $D_{2}$, associated with decaying mode, coming from the homogeneous solution, and it also involves $C_{1}$ arising from the inhomogeneous part. Since the leading order perturbation involves a growing mode, and we are interested in the first order correction to the TLN, it follows that it will be proportional to the ratio $(D_{2}/C_{1})$. This is because the TLN can be extracted from the coefficients of the $\order(r^2)$ and $\order(r^{-3})$ components, both of which are combinations of $C_{1}$ and $D_{2}$. This ratio can then be determined by imposing regularity at the horizon. With the appropriate numerical factors defined in \cite{Cardoso:2017cfl}, modulo the normalization with respect to the characteristic length scale of the problem, we obtain, 
\begin{equation}
k^{\rm E}_{l=2}=\frac{2M_{\rm DM}r_{s}^{4}\left[1+6\log(r_s/R)\right]}{R^{5}}\,.
\end{equation}
For the choice of the characteristic scale $R=r_{s}$, the TLN scales as $(M_{\rm DM}/r_{s})$, and hence is a linear quantity in $\epsilon$. This is consistent with our findings in the axial sector.

\subsubsection{Polar TLNs: Relativistic Schwarzschild-Hernquist profile}

Having described the TLN of a BH immersed in a non-relativistic Hernquist DM profile, we will now present the results for the relativistic Schwarzschild-Hernquist profile. As in the axial sector, here also the zeroth order perturbation $H^{(0)}$ satisfies the polar perturbation equation for the Schwarzschild spacetime, while the first order term satisfies a complicated equation. One can solve this equation in the near zone with the boundary conditions, $H^{(1)}(4\mbh)=0$ and $\partial_{r}H^{(1)}(4\mbh)=0$, and also solve the equation in the far zone in terms of an appropriate power series. Matching of the solution and its derivative between the near and the far zone uniquely fixes the arbitrary constants and, after setting $R=M_\text{BH}$ yields the following TLN for the polar $\ell=2$ mode, 

\begin{equation}
k^{\rm E}_{\ell=2}=645.9\epsilon_R\,.
\end{equation}
Note that, even though the axial and the polar TLNs have the same scaling they have different signs and magnitude, while the axial TLNs are negative, the polar TLNs are positive. We checked the value of the TLN is stable upon changing the matching radius, which validates our numerical method.

\section{Discussion}

Geometry of a black hole immersed in a dark matter halo and its response as well as relaxation to external perturbation will help in deciphering the presence of a dark matter environment, as well as establishing the density profile of the dark matter distribution from the GW measurements. The key feature of the dark matter distribution is its anisotropic nature: the radial pressure is negligible (assuming the accretion rate to be small compared to any other time scale in the problem), while the tangential pressure is non-zero. Finding out such a geometry started from Ref. \cite{Cardoso:2021wlq}, even though the dark matter density profile considered there was non-relativistic. Since the dark matter particles probe the region close to the horizon, it is important to study the relativistic generalization of the same, which was achieved in \cite{Speeney:2022ryg}. However, the fall-off behaviour for the dark matter density profile was inadequate, leading to serious issues, e.g., infinite energy.
In this work, we have derived, starting from the non-relativistic Hernquist dark matter profile, the relativistic profile for the dark matter distribution, using the approach of Ref. \cite{Sadeghian:2013laa}, and have referred to it as the Schwarzschild-Hernquist profile. The relativistic dark matter density profile, so derived, vanishes at $r=4\mbh$, has a sharp spike located at $r\approx 6\mbh$, and has the appropriate fall-off behaviour at large distances. It is worth pointing out that the peak of the relativistic dark matter spike is several orders of magnitude higher compared to the non-relativistic scenario. All of these features agree with the work \cite{Speeney:2024mas}, which was derived independently, and which nearly numerically agrees with our model.

Our construction of the relativistic dark matter profile brings an intriguing result involving the existence of multiple light rings. As is well known, existence of multiple light rings require the existence of stable and trapped photon orbits, which will eventually lead to instabilities of linear perturbations. Following the recent construction of geometries, which are solutions to the Einstein's equations with matter fields violating energy conditions, and having multiple light rings \cite{Guo:2022ghl}, we have also investigated the existence of such solutions within our phase space supported by an anisotropic dark matter distribution. We found that, indeed, there exist matter configurations that obey the dominant energy condition, while they admit multiple light rings, or, more precisely, light rings and anti-light rings in the sense of \cite{Cvetic:2016bxi}. Such matter configurations span a small region of parameter space, which we consider unphysical due to the very high matter densities involved, but we cannot rule them out just from known energy conditions. This suggests that realistic matter configurations do not have multiple light rings, but also that new energy conditions might exist that rule out such configurations, in line with independent work  \cite{Barcelo:2002bv,Martin-Moruno:2017exc} .

Besides pointing out various properties of the relativistic dark matter distribution, we have provided the complete formulation of linear perturbations of the coupled system of a dark matter profile supported by anisotropic matter and a central black hole, for both the axial and the polar sector. In comparison with the existing literature, we have considered an arbitrary radial pressure, which is independent from the pressure on the tangential directions. We wrote our final set of equations in readable and reproducible format on a \href{https://github.com/gcompere/Tidal-Love-numbers-and-quasi-normal-modes-in-the-Schwarzschild-Hernquist-dark-matter-halo}{Github} repository. During our analysis, we have invalidated the claim of \cite{Liu:2022csl} that a gauge-invariant third order master Zerilli-type equation exists for such perturbations. We also noted some differences with the earlier written equations in the literature \cite{Cardoso:2021wlq}: within the axial sector we have obtained two different equations, both in the dynamic as well as in the static situation, depending on whether we set the perturbation of contravariant or, covariant components of the fluid four-velocity to zero. Physically, these conditions correspond to distinct assumptions on the nature of the fluid perturbations. We have cross-checked our equations with two independent Mathematica codes, one based on the \href{www.xact.es}{xAct package} and another one based on the \href{http://www.inp.demokritos.gr/~sbonano/RGTC/}{RGtensor package}. We furthermore cross-checked the compatibility of the matter and the Einstein's equations. In addition, we have also discussed the static limit and the equations important for assessing the tidal Love numbers of the coupled BH-dark matter system. Thus our results have a much broader applicability: they hold for linear perturbation analysis of any anisotropic system, including anisotropic neutron stars \cite{Mondal:2023wwo}.  

In the dynamical sector, we have solved the master Regge-Wheeler-like equation for axial perturbations, with ingoing boundary condition at the horizon and outgoing boundary condition asymptotically. Our results demonstrate a key difference between the non-relativistic and relativistic dark matter distribution: for the non-relativistic case, the fractional change in the quasi-normal mode frequencies, as compared to the Schwarzschild black hole, scales linearly with $(M_{\rm DM}/r_{\rm s})$; while for the relativistic profile, the fractional change in the quasi-normal mode frequencies scale as $(M_{\rm DM}/r_{\rm s})^{3/2}$, with a mass dependent pre-factor. We therefore found a striking signature for a relativistic dark matter profile. Since our result is based on the Hernquist distribution, it would be interesting in the future to investigate how general are these scaling laws for more generic matter profiles. Moreover, there are indeed differences between the quasi-normal mode frequencies computed using either choice of matter perturbations (setting either the covariant or contravariant components of the perturbation of the fluid velocity to zero), but the difference is much smaller as compared to the quasi-normal mode frequencies themselves. The fractional change in the quasi-normal mode frequencies, as reported in this work have been numerically computed for dark matter profiles with unrealistic values, e.g., $M_{\rm DM}/\mbh=100$, for numerical stability reasons. For astrophysical scenarios with $M_{\rm DM}/\mbh=\mathcal{O}(10^{6})$, the fractional changes in the quasi-normal mode frequencies will be $\mathcal{O}(10^{-4})$, making it very difficult to distinguish between the Schwarzschild and the Schwarzschild-Hernquist black hole. 

Besides the quasi-normal mode frequencies, we have also derived both the axial and the polar tidal Love numbers for a black hole immersed in a relativistic as well as non-relativistic dark matter halo for realistic values of the dark matter profile. Our results agree with those in Ref. \cite{Cardoso:2021wlq}, for the non-relativistic Hernquist dark matter profile, except for the choice of the characteristic radius. Ref. \cite{Cardoso:2021wlq} takes this characteristic radius to be related to the mass of the dark matter halo, leading to a Love number which scales according to the ratio of dark matter total mass and breaking radius of the halo inconsistently with the perturbative approach considered. Instead, we have chosen this characteristic radius to be the breaking radius of the dark matter halo, leading to a tidal Love number which scales consistently with the perturbative scheme. Our results suggest that for realistic dark matter distribution, the tidal Love number for a non-relativistic distribution is $\mathcal{O}(10^{-4})$ while it suppressed to $\mathcal{O}(10^{-5})$ for a relativistic distribution with a central black hole of a million solar masses. Hence it is challenging to see the effect of the dark matter halo through the tidal Love number measurements, alike the situation for the quasi-normal modes. In addition, we have shown that the relative difference between the choice of matter perturbations lead to leading order $\mathcal{O}(1)$ changes of the tidal Love numbers. It is therefore crucial to model the fluid perturbations in detail in order to obtain meaningful predictions for the tidal Love numbers.   

In the process we have developed our own numerical methods 
to compute the tidal Love numbers for a black hole within the relativistic halo, with the result being highly numerically stable. In this numerical approach, we formulate an asymptotic analytic expansion of the perturbation variable at large radius, and at the same time solve for the relevant differential equation at arbitrary precision to low radius, so that we obtain the solution at small radius to high numerical precision. The tidal Love numbers are then obtained by matching the asymptotic solution with the near-zone solution at an intermediate radius. The resulting tidal Love numbers are found to be independent of the matching radius, as long as it belongs to a suitable range dictated by controlled numerical errors. This provides a robust scheme for deriving the tidal Love numbers for physically realistic values of the dark matter distribution. 

All in all, our results conclusively suggests that for realistic values of the dark matter distribution, both the quasi-normal modes, as well as the tidal Love numbers are modified by terms $\mathcal{O}(10^{-4})$ or lower. This suggests that the tidal Love numbers and the quasi-normal modes may not be able to provide a smoking gun signature for the existence of dark matter environments using the currently planned detectors. Rather the dynamical friction arising out of an extreme-mass-ratio inspiral within the dark matter environment can lead to significant dephasing of the GW signal and appears to be the proper avenue for exploring properties of dark matter environments \cite{Speeney:2022ryg, Speeney:2024mas, Clough:2021qlv, Mitra:2023sny, Traykova:2023qyv}. 

\section*{Acknowledgements}

We thank T. Hertog for related discussions. Research of SC is supported by MATRICS (MTR/2023/000049) and Core Research (CRG/2023/000934) Grants from SERB, ANRF, Government of India. SC also acknowledges the warm hospitality at the Albert-Einstein Institute, Potsdam, where a part of this work was done. The visit was supported by a Max-Planck-India mobility grant. LM is currently a PhD fellow at the Research Foundation - Flanders (FWO grant 1186024N). LM acknowledges funding from ESA Prodex project ’LISA EMRI/IMRAC waveform modelling’ PEA 4000131558, which financed earlier stages of this research. GC is Research Director of the F.R.S.-FNRS.

\appendix
\labelformat{section}{Appendix #1} 
\labelformat{subsection}{Appendix #1}
\section{The perturbation equations in vacuum Einstein gravity}
\label{app:Schw}

Let us now work out the case of polar gravitational perturbation in the context of vacuum Einstein gravity. In this case, the perturbed vacuum Einstein's equations become, $\delta G_{\mu \nu}=0=\delta R_{\mu \nu}$. we obtain, $H_{0}=H_{2}\equiv H(r)$. With this choice, the following differential equations can be obtained for the perturbations associated with the vacuum Einstein gravity,
\begin{align}
K'&+\left(\frac{r-3M}{r(r-2M)}\right)K-\frac{1}{r}H+\left(\frac{\ell(\ell+1)}{2i\omega r^{2}}\right)H_{1}=0~,
\label{zerilli3r}
\\
H_{1}'&+\left(\frac{2M}{r(r-2M)}\right)H_{1}+\frac{i\omega r}{r-2M}K+\frac{i\omega r}{r-2M}H=0~,
\label{zerilli2r}
\\
H'&+\left(\frac{r-3M}{r(r-2M)}\right)K+\left[\frac{i\omega r}{r-2M}+\frac{\ell(\ell+1)}{2i\omega r^{2}} \right]H_{1}-\frac{r-4M}{r(r-2M)}H=0~.
\label{zerilli4br}
\end{align}
In addition, we have the following algebraic relation between these three perturbation quantities,
\begin{align}\label{zerilli5r}
\Bigg[\frac{2M(r-3M)}{r(r-2M)}&-\frac{2\omega^{2}r^{3}}{r-2M}+(\ell-1)(\ell+2)\Bigg]K+\left[2i\omega r+\frac{\ell(\ell+1)M}{i\omega r^{2}} \right]H_{1}
-\left[\frac{6M}{r}+(\ell+2)(\ell-1)\right]H=0~.
\end{align}
However, we have two additional equations, 
\begin{align}
K''&+\left(\frac{3r-5M}{r(r-2M)}\right)K'-\frac{(\ell-1)(\ell+2)}{2r(r-2M)}K-\frac{1}{r}H'-\left(\frac{\ell(\ell+1)+2}{2r(r-2M)}\right)H=0~,
\label{zerilli1r}
\\
K''-H''&+\left[\frac{2(r-M)}{r(r-2M)}\right]K'-\left[\frac{2r}{r(r-2M)}\right]H'
\nonumber
\\
&+\left(\frac{\omega^{2}r^{2}}{(r-2M)^{2}}\right)H+\left(\frac{\omega^{2}r^{2}}{(r-2M)^{2}}\right)K-\left(\frac{2i\omega r}{r-2M}\right)H_{1}'-\left[\frac{2i\omega (r-M)}{(r-2M)^{2}}\right]H_{1}=0~.
\label{zerilli6br}
\end{align}
The first equation from the above set can be derived in the following manner --- (a) one first takes a derivative of Eq. \eqref{zerilli3r}, (b) then $H_{1}'$ is replaced in terms of other perturbation quantities, using Eq. \eqref{zerilli2r}, (c) and finally one uses Eq. \eqref{zerilli3r} to replace any remaining $H_{1}$. The derivation of Eq. \eqref{zerilli6br} follows the below mentioned route: (i) subtracting Eq. \eqref{zerilli4br} from Eq. \eqref{zerilli3r}, (ii) taking a derivative of the resulting expression, (iii) adding to it, the difference between Eq. \eqref{zerilli3r} and Eq. \eqref{zerilli4br}, multiplied by $\{2(r-M)/r(r-2M)\}$ and finally (iv) using the expression for $H_{1}'$ from Eq. \eqref{zerilli2r}. Thus, neither Eq. \eqref{zerilli1r}, nor Eq. \eqref{zerilli6br} are independent equations. In particular, given Eq. \eqref{zerilli5r}, along with any two of the three equations from Eq. \eqref{zerilli3r} to Eq. \eqref{zerilli4br}, the other can be derived.  

Let us now try to derive the master equation. For that purpose, let us solve for $H(r)$ from Eq. \eqref{zerilli5r} and then substitute the same in the differential equations for $K$, in Eq.  \eqref{zerilli3r}, yielding, 
\begin{align}
K'&+\left(\frac{r-3M}{r(r-2M)}\right)K+\left(\frac{\gamma_{\ell}+2}{2i\omega r^{2}}\right)H_{1}
\nonumber
\\
&-\frac{1}{\left(6M+\gamma_{\ell}r\right)}\left[\Bigg(\frac{2M(r-3M)+r(r-2M)\gamma_{\ell}-2\omega^{2}r^{4}}{r(r-2M)}\Bigg)K
+\left(\frac{-2\omega^{2}r^{3}+\left(\gamma_{\ell}+2\right)M}{i\omega r^{2}} \right)H_{1}\right]=0~.
\end{align}
where, we have defined $\gamma_{\ell}=(\ell+2)(\ell-1)$. Further simplification and rearrangement of the terms yield,
\begin{align}
K'=\left[\alpha_{0}(r)+\alpha_{2}(r)\omega^{2}\right]K+\left[\beta_{0}(r)+\beta_{2}(r)\omega^{2}\right]\left(\frac{H_{1}}{\omega}\right),
\label{Kprimelin}
\end{align}
with the following definitions for the known functions $\alpha_{0}(r)$, $\alpha_{2}(r)$, $\beta_{0}(r)$ and $\beta_{2}(r)$, respectively, as,
\begin{align}
\alpha_{0}&\equiv -\frac{-\gamma_{\ell}Mr+4M(r-3M)}{r\left(6M+\gamma_{\ell}r\right)(r-2M)}~,
\qquad
\alpha_{2}\equiv -\frac{2r^{3}}{\left(6M+\gamma_{\ell}r\right)(r-2M)}~,
\\
\beta_{0}&\equiv -\frac{\left(\gamma_{\ell}+2\right)\left(4M+\gamma_{\ell}r\right)}{2ir^{2}\left(6M+\gamma_{\ell}r\right)}~,
\qquad
\beta_{2}\equiv \frac{2ir}{\left(6M+\gamma_{\ell}r\right)}~.
\end{align}
On the other hand, the differential equation for $H_{1}$, as in Eq.  \eqref{zerilli2r}, becomes,
\begin{align}
H_{1}'&+\left(\frac{2M}{r(r-2M)}\right)H_{1}+\frac{i\omega r}{r-2M}K
\nonumber
\\
&+\frac{i\omega r^{2}}{\left(r-2M\right)\left(6M+\gamma_{\ell}r\right)}\left[\Bigg(\frac{2M(r-3M)+r(r-2M)\gamma_{\ell}-2\omega^{2}r^{4}}{r(r-2M)}\Bigg)K
+\left(\frac{-2\omega^{2}r^{3}+\left(\gamma_{\ell}+2\right)M}{i\omega r^{2}} \right)H_{1}\right]=0~.
\end{align}
This can also be expressed as,
\begin{align}
\frac{H_{1}'}{\omega}=\left[\kappa_{0}(r)+\kappa_{2}(r)\omega^{2}\right]K+\left[\delta_{0}(r)+\delta_{2}(r)\omega^{2}\right]\left(\frac{H_{1}}{\omega}\right)
\label{Hprimelin}
\end{align}
where, the coefficients $\kappa_{0}$, $\kappa_{2}$, $\delta_{0}$ and $\delta_{2}$ becomes,
\begin{align}
\kappa_{0}&\equiv -\frac{ir\left\{2\left(r-2M\right)\left(3M+\gamma_{\ell}r\right)+2M(r-3M)\right\}}{\left(r-2M\right)^{2}\left(6M+\gamma_{\ell}r\right)}~,
\qquad
\kappa_{2}\equiv \frac{2ir^{5}}{\left(r-2M\right)^{2}\left(6M+\gamma_{\ell}r\right)}~,
\\
\delta_{0}&\equiv -\frac{2M\left(6M+\gamma_{\ell}r\right)+\left(\gamma_{\ell}+2\right)Mr}{r\left(r-2M\right)\left(6M+\gamma_{\ell}r\right)}~,
\qquad
\delta_{2}\equiv \frac{2r^{3}}{\left(r-2M\right)\left(6M+\gamma_{\ell}r\right)}~.
\end{align}
Let us introduce a new function $\Psi_{\rm Z}$, such that,
\begin{align}
K&=\alpha(r)\Psi_{\rm Z}+\beta(r)\dfrac{d\Psi_{\rm Z}}{dr_{*}}~;
\label{zerillitest1}
\\
\frac{H_{1}}{\omega}&=\kappa(r)\Psi_{\rm Z}+\delta(r)\dfrac{d\Psi_{\rm Z}}{dr_{*}}~.
\label{zerillitest2}
\end{align}
where, the tortoise coordinate $r_{*}$ is defined as, $(dr/dr_{*})=(r-2M)/r$. Further, we demand, 
\begin{align}
\dfrac{d^{2}\Psi_{\rm Z}}{dr_{*}^{2}}=\left[V_{\rm Z}-\omega^{2}\right]\Psi_{\rm Z}~.
\end{align}
Then, taking derivative of Eq. \eqref{zerillitest1} with respect to the radial coordinate, we obtain,
\begin{align}
K'&=\alpha'\Psi_{\rm Z}+\left[\beta'+\alpha\left(\frac{r}{r-2M}\right)\right]\dfrac{d\Psi_{\rm Z}}{dr_{*}}+\beta\left(\frac{r}{r-2M}\right)\dfrac{d^{2}\Psi_{\rm Z}}{dr_{*}^{2}}
\nonumber
\\
&=\left[\alpha'+\beta\left(\frac{r}{r-2M}\right)\left(V_{\rm Z}-\omega^{2}\right) \right]\Psi_{\rm Z}+\left[\beta'+\alpha\left(\frac{r}{r-2M}\right)\right]\dfrac{d\Psi_{\rm Z}}{dr_{*}}.
\label{zerillitest2sc}
\end{align}
While from Eq. \eqref{Kprimelin}, and then using Eqs. \eqref{zerillitest1} and \eqref{zerillitest2sc}, we obtain, 
\begin{align}
K'&=\left[\alpha_{0}+\alpha_{2}\omega^{2}\right]\left(\alpha\Psi_{\rm Z}+\beta\dfrac{d\Psi_{\rm Z}}{dr_{*}} \right)+\left[\beta_{0}+\beta_{2}\omega^{2}\right]\left(\kappa\Psi_{\rm Z}+\delta\dfrac{d\Psi_{\rm Z}}{dr_{*}}\right)
\nonumber
\\
&=\left[\alpha\left(\alpha_{0}+\alpha_{2}\omega^{2}\right)+\kappa\left(\beta_{0}+\beta_{2}\omega^{2}\right)\right]\Psi_{\rm Z}
+\left[\beta\left(\alpha_{0}+\alpha_{2}\omega^{2}\right)+\delta\left(\beta_{0}+\beta_{2}\omega^{2}\right)\right]\dfrac{d\Psi_{\rm Z}}{dr_{*}}.
\end{align}
Comparing this with Eq. \eqref{zerillitest2sc}, we finally obtain,
\begin{align}
\alpha\left(\alpha_{0}+\alpha_{2}\omega^{2}\right)+\kappa\left(\beta_{0}+\beta_{2}\omega^{2}\right)&=\alpha'+\beta\left(\frac{r}{r-2M}\right)\left(V_{\rm Z}-\omega^{2}\right),
\\
\beta\left(\alpha_{0}+\alpha_{2}\omega^{2}\right)+\delta\left(\beta_{0}+\beta_{2}\omega^{2}\right)&=\beta'+\alpha\left(\frac{r}{r-2M}\right).
\end{align}
Thus equating coefficients of $\omega^{2}$ and terms independent of $\omega$, on both sides of the above expressions, we obtain,
\begin{align}
\beta\alpha_{2}+\delta\beta_{2}=0,
\\
\beta\alpha_{0}+\delta\beta_{0}=\beta'+\alpha\left(\frac{r}{r-2M}\right),
\\
\alpha\alpha_{0}+\kappa\beta_{0}=\alpha'+\beta\left(\frac{r}{r-2M}\right)V_{\rm Z},
\\
\alpha\alpha_{2}+\kappa\beta_{2}=-\beta\left(\frac{r}{r-2M}\right).
\end{align}
Choosing $\beta=1$, we get,
\begin{align}
\delta&=-\frac{\alpha_{2}}{\beta_{2}}=\frac{-ir^{2}}{r-2M}~,
\\
\alpha&=\left(\frac{r-2M}{r}\right)\left(\alpha_{0}+\delta\beta_{0}\right)= \frac{\gamma_{\ell}\left(\gamma_{\ell}+2\right)r^{2}+6\gamma_{\ell}Mr+24M^{2}}{2r^{2}\left(6M+\gamma_{\ell}r\right)},
\\
\kappa&=\alpha\delta-\frac{1}{\beta_{2}}\left(\frac{r}{r-2M}\right)=i\left[\frac{-\gamma_{\ell}r^{2}+3\gamma_{\ell}Mr+6M^{2}}{\left(6M+\gamma_{\ell}r\right)(r-2M)} \right].
\end{align}
Thus finally,
\begin{align}
V_{\rm Z}&=\left(\frac{r-2M}{r}\right)\left[\alpha\alpha_{0}+\kappa\beta_{0}-\alpha' \right]
\nonumber
\\
&=\left(\frac{r-2M}{r}\right)\Bigg[\frac{\gamma_{\ell}^{2}\left(\gamma_{\ell}+2\right)r^{3}+6M\gamma_{\ell}^{2}r^{2}+36M^{2}\gamma_{\ell}r+72M^{3}}{r^{3}\left(6M+\gamma_{\ell}r\right)^{2}}\Bigg]. 
\end{align}
This is the Zerilli potential, associated with polar perturbation of the Schwarzschild spacetime.

\bibliographystyle{utphys1}
\bibliography{bibliography}


\end{document}